\documentclass[preprintnumbers,amsmath,amssymb]{revtex4}

\usepackage{graphicx}
\usepackage{dcolumn}
\usepackage{bm}
\usepackage{array}

\usepackage{epstopdf}

\def\ket{\rangle}
\def\bra{\langle}

\begin{document}

\title{Density of states and quantum phase transition in the thermodynamic limit of the Mermin central-spin model}

\author{Savannah Sterling Garmon}
 \email{sgarmon@chem.utoronto.ca}
\affiliation{
Chemical Physics Theory Group,
Department of Chemistry and Center for Quantum Information and Quantum Control,
University of Toronto,
80 St. George Street,
Toronto, Ontario,
Canada M5S 3H6}

\author{Pedro Ribeiro}
\email{ribeiro@cfif.ist.utl.pt}
\affiliation{CFIF, Instituto Superior T\'{e}cnico, Universidade
T\'{e}cnica de Lisboa, Av. Rovisco Pais, 1049-001 Lisboa, Portugal}


\author{R\'emy Mosseri}
\email{remy.mosseri@upmc.fr}
\affiliation{%
 Laboratoire de Physique Th\'eorique de la Mati\`ere Condens\'ee, 
 CNRS UMR 7600, 
 Universit\'e Pierre et Marie Curie,
 4 Place Jussieu, 75252 Paris Cedex 05, France}

\date{\today}

\begin{abstract}
We apply a spin-coherent states formalism to study the central-spin model with monochromatic bath and symmetric coupling (the Mermin model); in particular, we derive analytic expressions for the integrated density of states in the thermodynamic limit when the number of bath spins is taken to infinity.  From the thermodynamic limit spectrum we show the phase diagram for the system can be divided into four regions, partitioned on the one hand into a symmetric (non-degenerate) phase or a broken symmetry (degenerate) phase, and on the other hand by the case of overlapping or non-overlapping energy surfaces.  The nature and position of singularities appearing in the energy surfaces change as one moves from region to region.  Our spin-coherent states formalism naturally leads us to the Majorana representation, which is useful to transform the 
Schr\"odinger
equation into a Ricatti-like form that can be solved in the thermodynamic limit to obtain closed-form expressions for the integrated density of states.  The energy surface singularities correspond with critical points in the density of states.  We then use our results to compute expectation values for the system that help to characterize the nature of the quantum phase transition between the symmetric and broken phases.
\end{abstract}

\maketitle

\section{introduction}

The study of a two-level system interacting with some form of environment or background has been the focus of much research \cite{cent_spin_review,spin_boson_review,UWeiss}.  Such studies have focused on important questions such as the onset of decoherence in nature as well as the transition from quantum to classical physics.  The former question takes on a concrete practical importance in the context of the emerging field of quantum information processing, in which case the two-level system could be used to simulate the time-evolution of a qubit \cite{Wubs2006}.  In this particular situation, the goal would be to maintain the coherence of the two-level system on some sufficiently long time scale in order to preserve the information embedded in the qubit until the next step in the computation process occurs.

While these studies have been considered in the context of a wide range of natural phenomena, interestingly it seems these models fall into two distinct classes based on the characterization of the environment \cite{cent_spin_review}.  
On the one hand, our physical picture could be seen as an environment composed of localized impurities, each with a small number of low energy states that can be modeled as a spin 1/2 system or a set of such spins.
In the case of the two-level central system coupled to such a `spin bath,' this is called the central-spin model.  A special case of this model will be the focus of the present study.  On the other hand we could also consider a two-level system coupled to a large number of background oscillators, which is the spin-boson model.

Here we focus on the so-called Mermin model, which is the special case of the central-spin model in which the background spins share a common frequency (the `monochromatic bath') and each interact symmetrically with the central spin.  This model was proposed by N. D. Mermin \cite{Mermin} as an illustration of the transition from a delocalized quantum state to a localized state (with broken spatial symmetry) in the classical limit.
Since then, the dynamical properties of the system across the phase transition have been studied \cite{Lev2000,Lev&Muth2001}, and the model has also been studied under the names of the finite Jaynes-Cummings Model (or su(2) Jaynes-Cummings model) \cite{Ell&Kov} and the spin star system \cite{Non-Markovian}.  In the latter case, the authors have applied various approximation techniques to the density operator evolution in order to demonstrate decoherence and the presence of non-Markovian effects.

In the present study, we will apply a spin coherent states \cite{Raddiffe,Arrechi,Klauder} based approach to the Mermin model in order to obtain the analytic form of the integrated density of states in each phase region of the model.  By introducing the spin coherent states, we will find that the wave function for the system is naturally transformed into the Majorana representation \cite{Majorana}.  This approach follows that which was recently applied to analyze the spectrum of the well-known Lipkin-Meshkov-Glick (LGM) model \cite{Ribeiro1,Ribeiro2}.  However, it is worth noting that this type of analysis has proved fruitful in other studies as well.  For example, a similar approach (also based on the spin coherent states) has previously been applied to the Mermin (or finite JCM) model \cite{Ell&Kov}, although with the aim of characterizing the system in a somewhat different manner, as we will briefly elaborate on in our concluding remarks.  As another example, a generalized SU($M$) coherent states approach has been recently applied in investigations of the $M$-site Bose Hubbard model \cite{Fried&Dirk1}.

The organization of our paper is as follows.  In Sec. \ref{SEC:model.ham} we write the Mermin model Hamiltonian and introduce the system representation in terms of the spin coherent states that transforms the wave function into the form of the Majorana polynomial.
The roots of this polynomial can then be used to characterize the system.  In Sec. \ref{SEC:energy.surface}, we apply a mean field theory analysis to find the classical energy sheets in the thermodynamic limit.  There are two such sheets, one each associated with the two available spin states of the central system.  We will further study the extrema of these energy sheets and find that their behavior serve to characterize the phase diagram.  It is well known that, due to spontaneous symmetry breaking, the phase diagram for the Mermin model can be divided into a symmetric phase (with non-degenerate ground state) and a (degenerate) broken symmetry phase.  We will see that these two regions can be further divided according to whether the two energy sheets are overlapping or non-overlapping.

We will then turn to the analysis of the time-independent Schr\"odinger equation in Sec. \ref{SEC:diffeq.thermo}.  
Using the coherent states formalism we show that the Schr\"odinger equation can be mapped into a first-order nonlinear differential equation in a Ricatti-like form.  By expanding these equations in powers of the inverse number of background spins $1/N$, we formulate a method by which we can determine an analytic form for the integrated density of states in the thermodynamic limit (infinite $N$).  
The results of these calculations are presented in Sec. \ref{SEC:anal.dos}, with some details of how the density of states expressions may be obtained provided in the Appendix.
Then in Sec. \ref{SEC:observables} we will make use of our results to compute the expectation values for several key quantities in the system.  In part, this will serve to help in the characterization of the quantum phase transition.  Finally in Sec. \ref{SEC:conclusion} we summarize our results and make some comments on future work.


\section{Hamiltonian and Majorana Representation}\label{SEC:model.ham}

The Mermin model describes the interaction of a two-level system, represented by the central spin or `small spin' 
$\boldsymbol{ \sigma}$, with a collection of $N$ background spins $\boldsymbol{\sigma}^i$.  Considering only the fully symmetric subspace, these background spins can be summed into one large `environment spin' 
${\bf S} = \sum_{i=1}^N \boldsymbol{ \sigma}^i / 2$ with maximal spin $S=N/2$.  Here the $\sigma$'s are the standard Pauli spin matrices, although we will introduce a different representation for the central spin below.  The interaction term will then be given by an anti-ferromagnetic coupling between the $x$ components of the central spin and the environment spin, according to
\begin{equation}
\mathcal H= \frac{ \omega_z}{2} \sigma_z  + \Omega \frac{S_z}{2S} + \gamma_x \sigma_x \frac{S_x}{2S}
\label{ham.mermin}
\end{equation}
with $\gamma_x > 0$.
Also notice that the second and third terms involving the collective spin have been re-scaled by the factor $1/2S$ in order to obtain a non-trivial phase diagram.  

Let us now introduce the spin coherent states basis that will lead us to the Majorana representation.  
First, we introduce the most familiar representation for this system in terms of the tensor product  $|S, M \ket  \otimes |s,m \ket $ of the environment spin basis $|S, M \ket$ and the small spin 
basis $|s,m \ket$ by writing a general state as $| {\bf \Psi} \ket 
 = \sum_{M,m} c_{M,m} ( |S, M \ket \otimes |s,m \ket )$.
Here $M = -S, \dots ,S$ and $m = -s, \dots , s$ where, for the system we have in mind, $S \gg s$.
Then the coherent states representation $| \alpha \ket$ in terms of the collective spin $S$ is given by
\begin{equation}
\bra \alpha | {\bf \Psi} \ket 
 = \sum_m  \sum_M c_{M,m} \bra \alpha | S,M \ket 
 |s, m \ket
\label{psi.coherent.vector}
\end{equation}
where $\alpha$ is a complex parameter; we now define
\begin{equation}
\psi_m(\alpha) \equiv \sum_M c_{M,m} \bra \alpha | S,M \ket = C \prod_{k=1}^d ( \alpha - \alpha_k^{(m)}  )
\label{psi.coherent},
\end{equation}
as the Majorana polynomial of order $d \le N$ with maximum order $N=2S$.  The coherent state 
$| \alpha \ket$ is given by  $| \alpha \ket = e^{\alpha S_+} | S, - S \ket$, which can be shown to be equivalent to
\begin{equation}
| \alpha \ket = \sum_{M=-S}^{S} \sqrt{ \frac{(2S)!}{(S+M)! (S-M)!} } \alpha^{S+M} | S, M \ket.
\label{alpha.expansion}
\end{equation}
The inner product of two spin coherent states is given by
\begin{equation}
\bra \alpha' | \alpha \ket  =  (1 + \bar{\alpha}' \alpha)^{2S}.
\label{inner.alpha}
\end{equation}
In the present representation, the environmental spin operators appearing above are given by
\begin{eqnarray}
S_z & = & - S + \alpha \partial_\alpha \nonumber \\
S_- & = & \partial_\alpha \label{S.ops.coherent.rep} \\
S_+ & = & 2 S \alpha - \alpha^2 \partial_\alpha \nonumber
\end{eqnarray}
along with the ordinary relations $S_\pm = S_x \pm i S_y$.  

Now we will choose to specify that $s = 1/2$ (Mermin's model).
Let us write our vector-valued wave function in the explicit Pauli representation as
\begin{equation}
{\bf \Psi}(\alpha) = \left(
\begin{array}{c}
      \psi_{1/2} (\alpha)  \\
      \psi_{-1/2}  (\alpha)
\end{array}
\right)
\equiv \left(
\begin{array}{c}
      \psi_{\uparrow} (\alpha)  \\
      \psi_{\downarrow}  (\alpha)
\end{array}
\right)
\label{state.vector.s1/2}
\end{equation}
with the functions $\psi_{\uparrow, \downarrow} (\alpha)$ given through Eq. (\ref{psi.coherent}).
Before proceeding with our analysis, we find that a change of basis for the small spin following the unitary transformation
\begin{equation}
\mathcal U = {1 \over \sqrt{2}}
\left[
\begin{array}{rr}
1 & 1 \\
1 & -1 
\end{array}
\right].
\label{unit.trans.sa.rep}
\end{equation}
will help simplify the analysis of the  time-independent Schr\"odinger equation in Sec. \ref{SEC:diffeq.thermo}.
Applying this transformation to Eq. (\ref{state.vector.s1/2}) we find that the wave vector in the `up, down' representation is transformed into the `symmetric, anti-symmetric' representation according to
\begin{equation}
\mathcal U^{\dagger}
\left[
\begin{array}{c}
\psi_{\uparrow}  (\alpha) \\
\psi_{\downarrow}  (\alpha)
\end{array} \right]
=  {1 \over \sqrt{2}}
\left[
\begin{array}{c}
\psi_{\uparrow} (\alpha) + \psi_{\downarrow}  (\alpha) \\
\psi_{\uparrow} (\alpha)  -  \psi_{\downarrow} (\alpha)
\end{array}  \right]
\equiv
\left[
\begin{array}{c}
\phi_s (\alpha) \\
\phi_a (\alpha)
\end{array}  \right]
\label{wavefcn.sa.basis}
\end{equation}
Note that in this basis, the $\bar{\sigma}_x  = \mathcal U^\dagger \sigma_x \mathcal U$ matrix (which includes the interaction term in Eq. (\ref{ham.mermin}) above) is diagonalized.
As a result, the Hamiltonian (\ref{ham.mermin}) takes the form
\begin{equation}
H = \left[
\begin{array}{cc}
\Omega \frac{S_z}{2S} + \gamma_x \frac{S_x}{2S}  	&   \frac{\omega_z}{2}    \\
\frac{\omega_z}{2}         						&  \Omega \frac{S_z}{2S} - \gamma_x \frac{S_x}{2S}
\end{array}  \right]
\label{eigen.eqns.matrix.sa}
\end{equation}
in which operator terms appear only on diagonal entries, 
which is an indication that the Schr\"odinger equation may be written in a more compact form in this basis.


\section{Classical Energy Surfaces and Phase Diagram}\label{SEC:energy.surface}

In this section we will introduce a mean field theory (MFT) approximation to find the two classical energy surfaces, one each associated with the two degrees of freedom for the central spin.  From the energy surfaces we can then determine the critical points in the spectrum, including the ground state.  
These are singular points such as saddle points that may occur inside the energy continuum or the max/min that form the edges of the continuum for a given sheet.
The ground state is given by the minimum on the lower energy surface.
The phase diagram will then be revealed by studying the critical points as a function of the Hamiltonian parameters.

\subsection{Mean Field Theory and Classical Energy Surfaces} \label{SEC:energy.surface.MFT}

In the context of the spin coherent states representation we introduce
a mean field theory approximation, such that  correlations
between the bath and the central spin factorize, as
${\bf \Psi} \sim {\bf \Psi}_{MFT} (\alpha) =  \mbox{\boldmath $\eta$}
(1 + \bar{\alpha} \alpha )^{2S}$ with $ \mbox{\boldmath $\eta$} =
(\eta_s,\eta_a)$ a constant complex vector.
With this factorized form we are in essence assuming that in the MFT each individual background spin will have its own distinguishable interaction with the central spin, and that these interactions will be symmetric.
We will further take the thermodynamic limit $N = 2S \rightarrow \infty$ to obtain our expressions for the classical energy surfaces below.

Let us make a brief comment on our approximation before proceeding.
Note that the thermodynamic limit in this context is equivalent to the classical limit in the sense that as $S \rightarrow \infty$ the large environment spin will behave as a classical spin, since
we have chosen to work in the symmetric subspace where the individual bath spins can be summed into a single giant spin.

Applying our mean field ansatz, we can obtain the classical energy surfaces by further taking the limit $N = 2S \rightarrow \infty$ as
\begin{equation}
\epsilon_{\pm} (\alpha, \bar{\alpha}) = \frac{\Omega (-1 + \bar{\alpha} \alpha ) \pm \sqrt{\omega^2 (1 + \bar{\alpha} \alpha)^2 + \gamma_x^2 (\alpha + \bar{\alpha})^2 } }
{2 (1 + \bar{\alpha} \alpha)}.
\label{clas.energy.surface}
\end{equation}
This quantity gives a mean field description of the energy spectrum of our model parameterized as a function of the complex pair $\alpha, \bar{\alpha}$.  The singular points in the spectrum, such as edge singularities and saddle-points, can be exactly determined as the critical points of $\epsilon_{\pm} (\alpha, \bar{\alpha})$, a calculation which is carried out in the following section.  We plot the classical energy surfaces in terms of $x = \textrm{Re} \ \alpha, y = \textrm{Im} \ \alpha$ for four choices of the system parameters in Fig. \ref{fig:energy_surface_plots}.

Note that alternatively, we could apply a classical spin description that is equally valid in the thermodynamic limit.  This classical spin description is obtained via the transformation
\begin{equation}
S_x = \frac{N}{2} \sin \theta \cos \phi, S_y = \frac{N}{2} \sin \theta \sin \phi, S_z = \frac{N}{2} \cos \theta.
\label{clas.spin.defn}
\end{equation}
Indeed, the quantum phase transition between the symmetric and broken phases has previously been studied in terms of this equivalent description \cite{Mermin,Lev2000,Lev&Muth2001}.  
A transformation between the two representations can be accomplished through $\alpha = \tan \frac{\theta}{2} e^{i \phi}$.
There are a handful of calculations that are in fact more simple in one representation or the other, although we will focus on the $\alpha, \bar{\alpha}$ description below.  To obtain a better physical intuition for these complex parameters, we note the following relationship to the classical spin components
\begin{eqnarray}
 \textrm{Re}(\alpha)&  = & \frac{\bra S_x \ket}{N/2 + \bra S_z \ket}  \label{xComp}  \\
 \textrm{Im}(\alpha) & =  & \frac{\bra S_y \ket}{N/2 + \bra S_z \ket}  \label{yComp} \\
 (\textrm{Re}(\alpha))^2 + (\textrm{Im}(\alpha))^2  
 		& =  & \frac{(N/2)^2 - \bra S_z \ket^2}{(N/2 + \bra S_z \ket)^2},     \label{zComp}
\end{eqnarray}
obtained from the above transformation and Eq. (\ref{clas.spin.defn}).  Hence, roughly speaking, the real part of $\alpha$ can be associated with the $x$ component of spin and the imaginary part with the $y$ component.

\subsection{Domain of critical points in the $\alpha, \bar{\alpha}$ plane}\label{SEC:energy.surface.MFT.critpts}

In order to characterize the system and obtain the phase diagram, we determine the critical points of the spectrum from $\partial_\alpha \epsilon_{\pm} = \partial_{\bar{\alpha}} \epsilon_{\pm} = 0$.  
The solution to these conditions are obtained as roots of a quadratic equation in $\alpha^2$, along with the condition
$\alpha = \bar{\alpha}$ (that is, $\alpha$ real).
The four explicit solutions of the quadratic are given by $\pm \alpha_+, \pm \alpha_-$, in which
\begin{equation}
\alpha_\pm = \sqrt{ \frac{
\gamma_x^4 + \Omega^2 \omega_z^2 + 2 \gamma_x^2 \Omega^2 \pm 
2  \gamma_x \Omega \sqrt{ (\Omega^2 + \gamma_x^2) ( \omega_z^2 + \gamma_x^2) }
}
{\gamma_x^4 - \Omega^2 \omega_z^2}
}.
\label{crit.alpha.4points}
\end{equation}
There are two other sets of critical points given by $\alpha = \bar{\alpha} = 0$ and $\alpha = \bar{\alpha} \rightarrow \infty$.

\begin{figure}
 \includegraphics[width=0.7\textwidth]{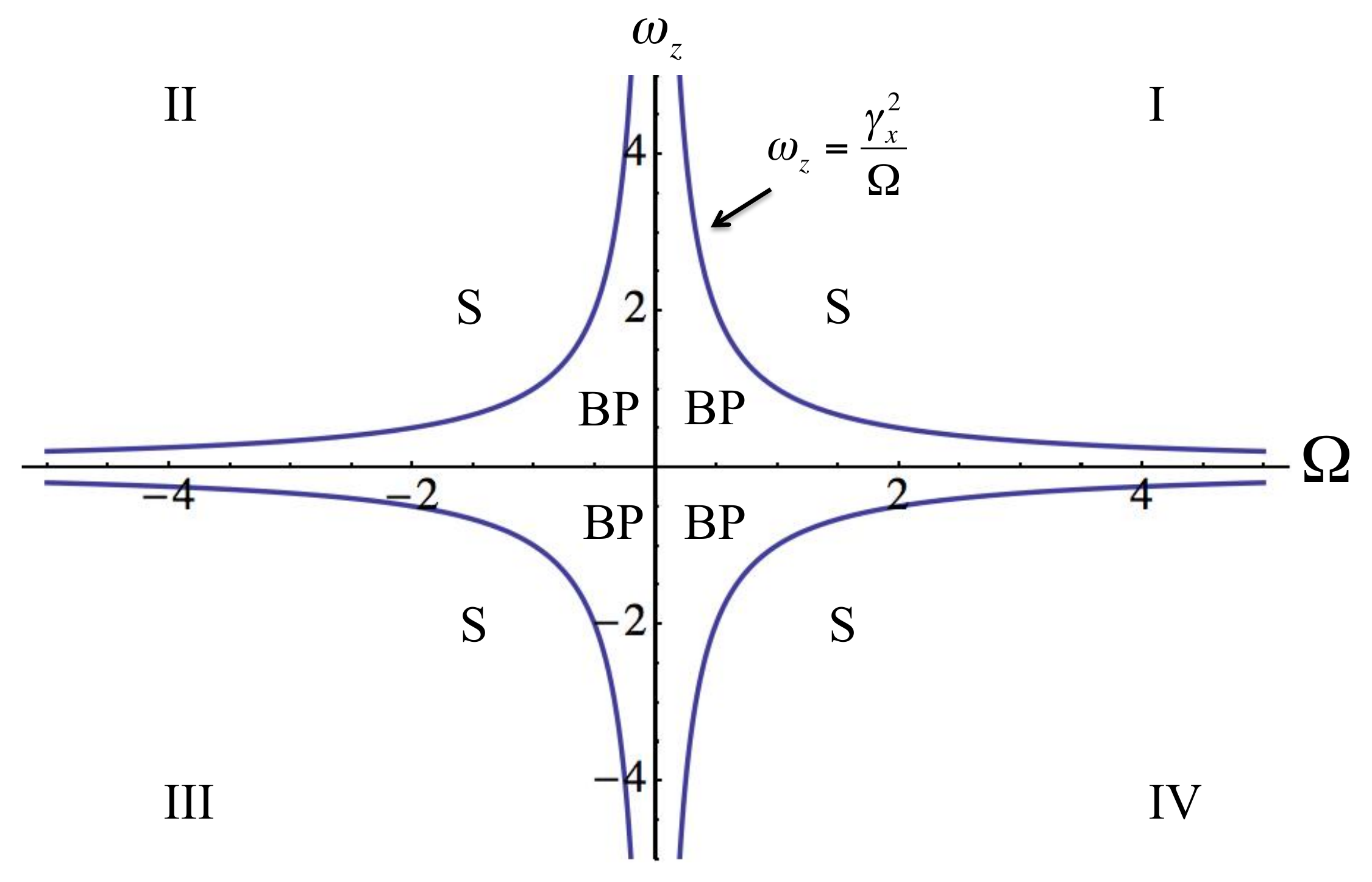}
 \caption{The four quadrants in parameter space for the Mermin model.  Here we have chosen the value $\gamma_x = 1$.  We show the line $|\omega_z| = \frac{\gamma_x^2}{|\Omega|}$ that denotes the quantum phase transition in each quadrant.   In each quadrant, we have also labeled the symmetric (S) and broken phase (BP) portion of the phase diagram.
  Each quadrant is equivalent according to the symmetry of the system, thus we will focus exclusively on Quadrant I as we proceed.  Note that we then further divide each quadrant into four regions in Fig.  \ref{fig:phase.regions} below.}
 \label{fig:phase.quads}
\end{figure}

We notice immediately that the denominator $\gamma_x^4 - \Omega^2 \omega_z^2 = (\gamma_x^2 - \Omega \omega_z)(\gamma_x^2 + \Omega \omega_z)$
of the critical points given in Eq. (\ref{crit.alpha.4points})  will vanish whenever
\begin{equation}
\gamma_x^2 = |\Omega| |\omega_z|
\label{cond.QPT}
\end{equation}
is satisfied.  Since $\Omega$ and $\omega_z$ may take either positive or negative values, this condition defines four symmetric curves in parameter space as shown in Fig. \ref{fig:phase.quads}.  Since the four quadrants given in this diagram are identical, we will focus throughout the paper on quadrant I where the above condition can simply be written $\gamma_x^2 = \Omega \omega_z$.
We will demonstrate below that this divergence indicates the presence of a quantum phase transition.

On one side of this transition ($\omega_z < \gamma_x^2 / \Omega$), these four critical points will all be real, with
$\pm \alpha_-$ 
forming two degenerate (global) minima on the lower sheet, while the two points $\pm \alpha_+$ will form degenerate (global) maxima in the upper sheet.  On the other side of the transition ($\omega_z > \gamma_x^2 / \Omega$), the points $\alpha_\pm$ become complex conjugates 
(violating $\alpha = \bar{\alpha}$) and there will be a single minimum (given by $\alpha = \bar{\alpha} = 0$) at the origin in the lower sheet, which will form the non-degenerate ground state in the symmetric phase.

Finally, notice from Eq. (\ref{xComp}) that the purely real points in Eq. (\ref{crit.alpha.4points}) are associated with a non-zero $x$ component to the spin.  Hence, the quantum phase transition is apparently induced by the coupling term in the Hamiltonian (which involves the $x$ component of the spins).  Writing the condition for the broken phase as $\gamma_x^2 > \Omega \omega_z$ makes clear that the symmetry breaking can be viewed as occurring when the energy scale of the coupling dominates over that of the spins.

\subsection{Extremal energy values, ground state degeneracy and quantum phase transition}\label{SEC:energy.surface.MFT.QPT}

Having obtained the critical values of $\alpha, \bar{\alpha}$, we will now find the explicit values of the extremal energies including the ground state.

We note that the critical points at the origin $\alpha = \bar{\alpha} = 0$ are critical points in \emph{both} sheets throughout parameter space.  
From Eq. (\ref{clas.energy.surface}) we immediately find
\begin{equation}
\epsilon_{\pm} (\alpha = 0, \bar{\alpha} = 0) = \frac{ - \Omega \pm \omega_z }{ 2 } \equiv \epsilon_\pm^0
\label{crit.energy.alpha.zero}
\end{equation}
The negative choice $\epsilon_-^0$ on the lower sheet acts as the (unique) ground state throughout the symmetric phase.  This point becomes a saddle point in the broken phase as we show below.
Meanwhile, for the pair of critical points at infinity $\alpha, \bar{\alpha} \rightarrow \infty$ we find
\begin{equation}
\epsilon_{\pm} (\alpha \rightarrow \infty, \bar{\alpha} \rightarrow \infty) = \frac{ \Omega \pm \omega_z }{ 2 }  \equiv \epsilon_{\pm}^{\infty}.
\label{crit.energy.alpha.infty}
\end{equation}

To verify the presence of the quantum phase transition, we construct the Hessian matrix that describes the qualitative behavior of a given critical point through the second derivatives of the classical energy surfaces. 
We construct the Hessian $\Theta_\pm (x, y)$ with $x =  \textrm{Re} \ \alpha, y = \textrm{Im} \ \alpha$
according to
\begin{equation}
\Theta_\pm (x, y) = \left[
\begin{array}{cc}
\partial_{x,x} \epsilon_\pm (x,y)  	& \partial_{x,y} \epsilon_\pm (x,y)  \\
\partial_{y,x} \epsilon_\pm (x,y) 	& \partial_{y,y} \epsilon_\pm (x,y) 
\end{array}
\right].
\label{hessian}
\end{equation}
We can now distinguish between maxima, minima and saddle points, which will have negative definite, positive definite, and indefinite Hessian, respectively.

\begin{figure*}
\hspace*{0.05\textwidth}
 \includegraphics[width=0.2\textwidth]{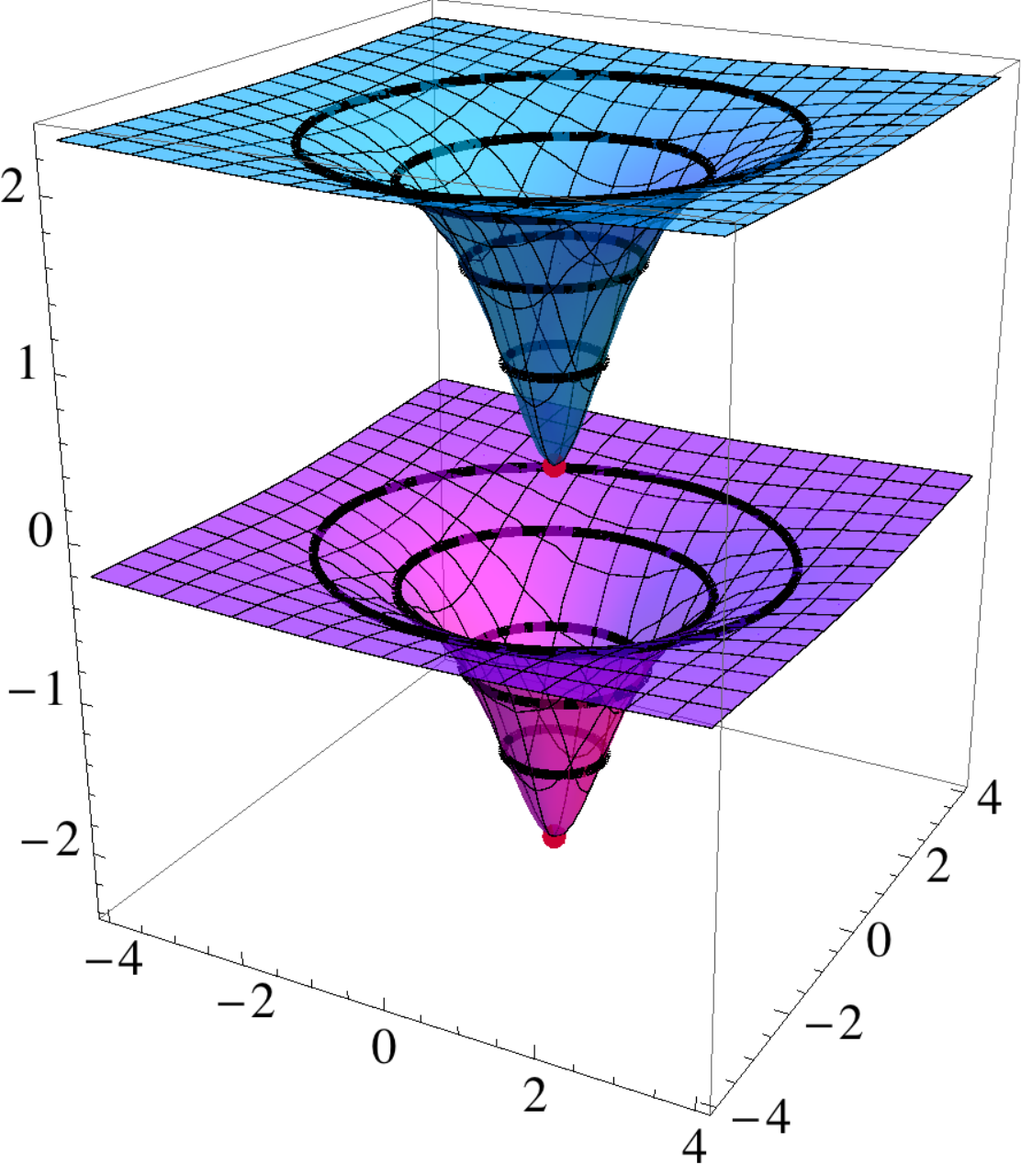}
\hfill
 \includegraphics[width=0.2\textwidth]{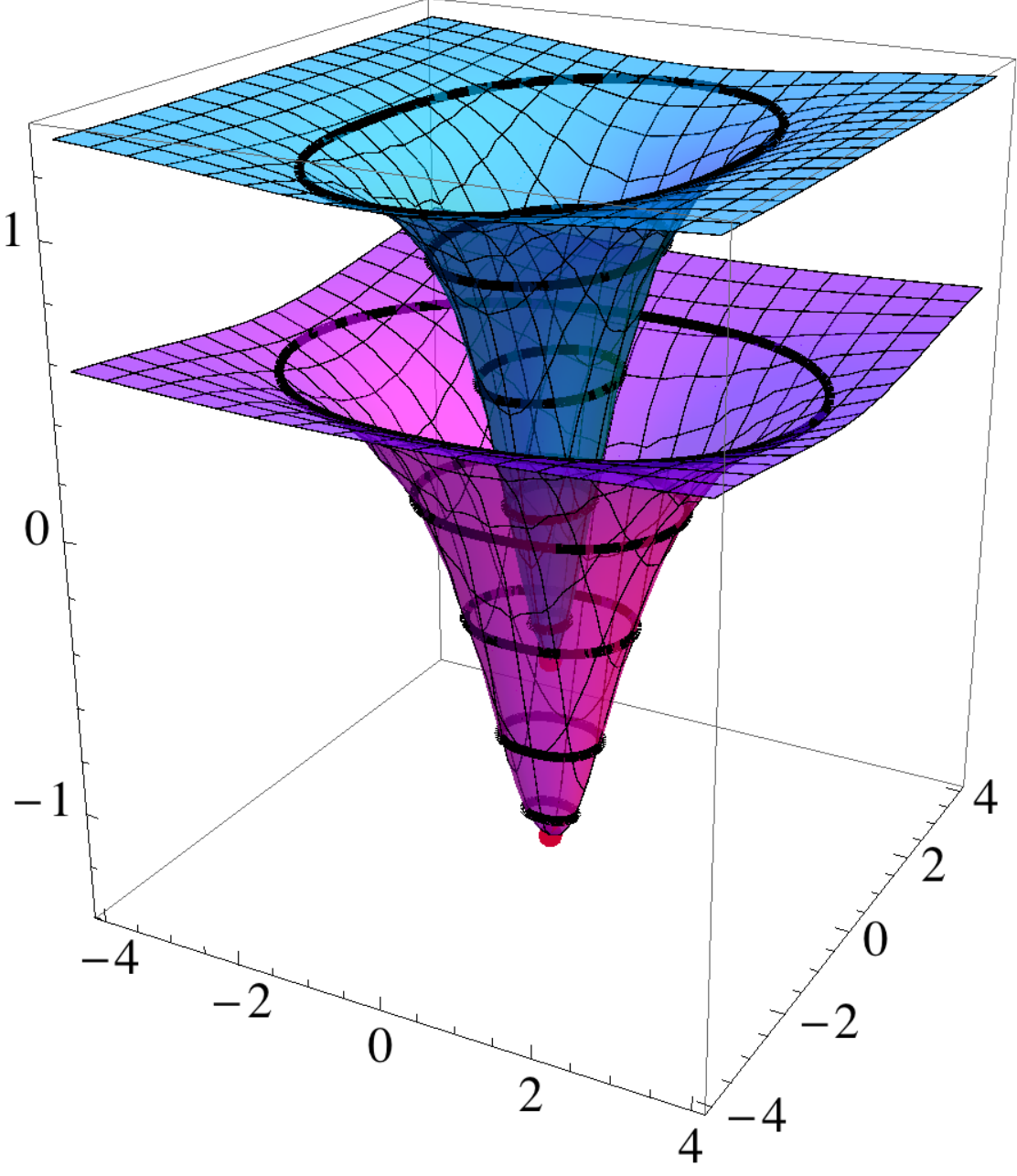}
\hfill
 \includegraphics[width=0.2\textwidth]{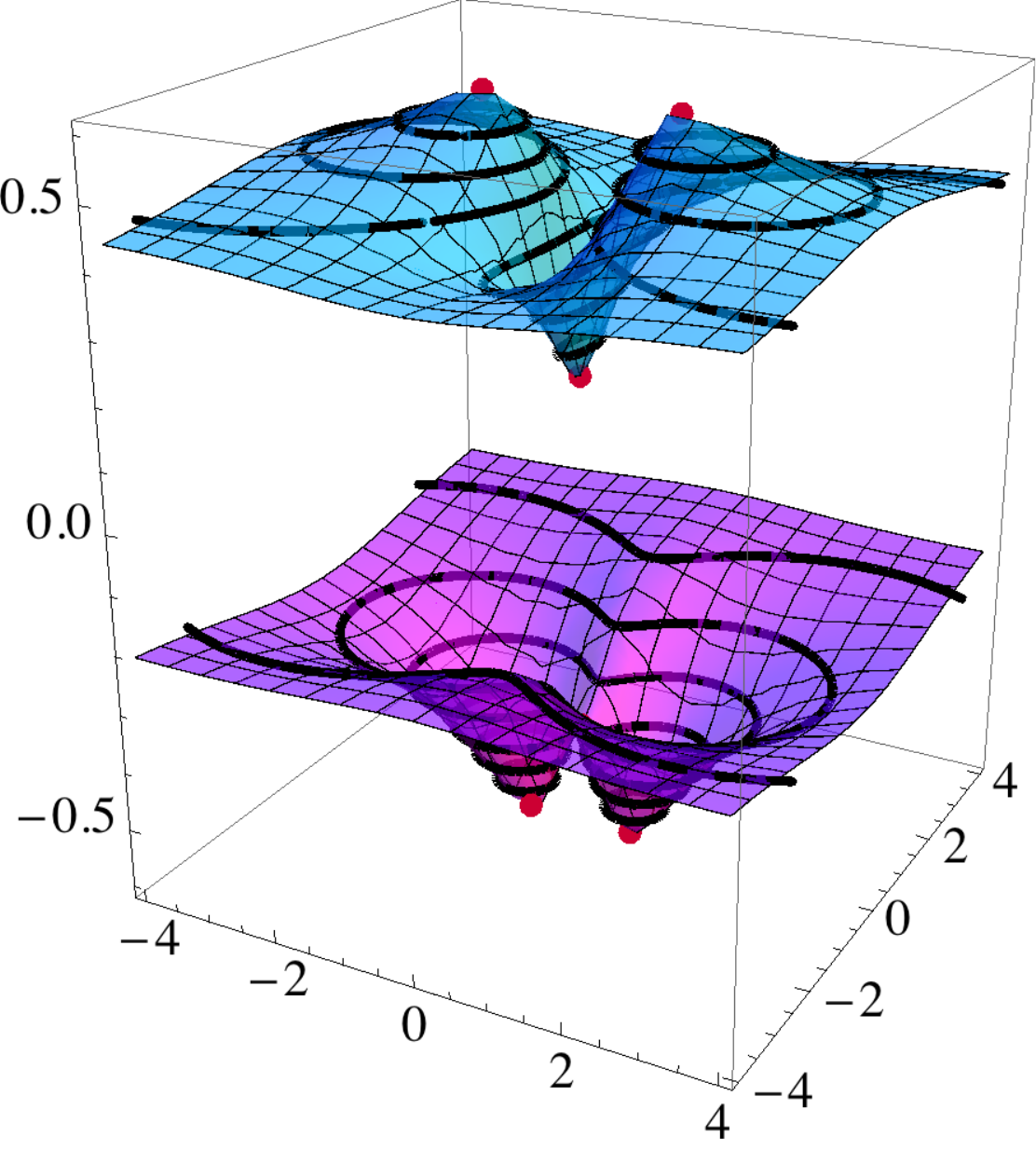}
\hfill
 \includegraphics[width=0.2\textwidth]{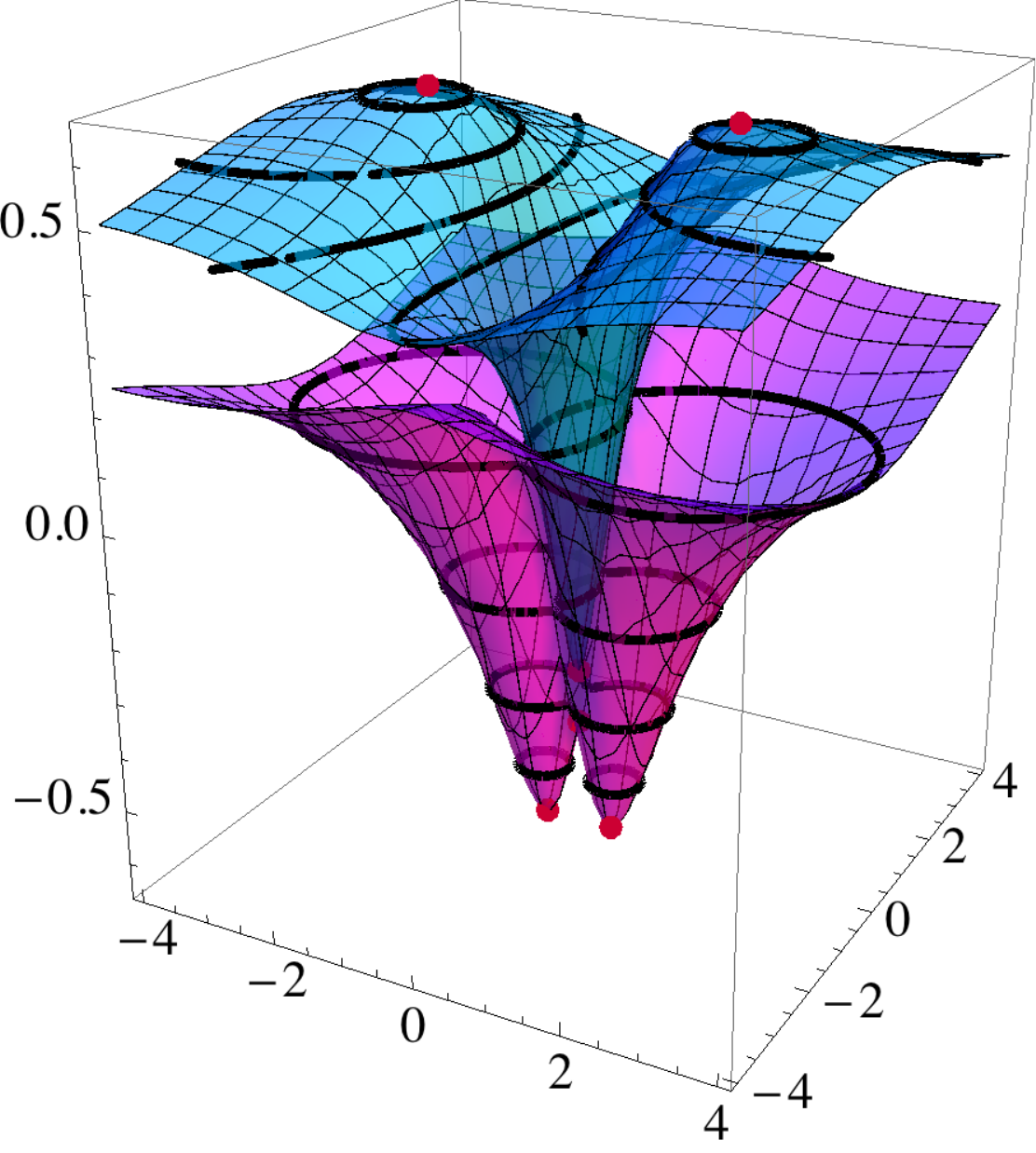}
 \hspace*{0.05\textwidth}
\\
\vspace*{-\baselineskip}
\hspace*{0.08\textwidth}(a)\hspace*{0.2\textwidth}(b)\hspace*{0.2\textwidth}
	(c)\hspace*{0.2\textwidth}(d)\hspace*{0.2\textwidth}
\\
\vspace*{\baselineskip}
 \caption{ (a) Region 1 (symmetric, non-overlapping), (b) Region 2 (symmetric, overlapping), (c) Region 3 (broken phase, non-overlapping), and (d) Region 4 (broken phase, overlapping).
Here we plot the two energy surfaces 
 as a function of $x = \textrm{Re} (\alpha)$ and $y = \textrm{Im} (\alpha)$ for each of the four phase regions in quadrant I.  We have used the values 
 (a) Region 1: $\Omega = 2.2$, $\omega_z = 2.5$,
 (b) Region 2: $\Omega = 2.0$, $\omega_z = 0.7$,
 (c) Region 3: $\Omega = 0.25$, $\omega_z = 0.6$, and
 (d) Region 4: $\Omega = 0.8$, $\omega_z = 0.1$
 along with the typical value $\gamma_x = 1$ in all cases.  
 Critical points are shown as red 
points and some level energy surfaces are drawn in black. }
 \label{fig:energy_surface_plots}
 \end{figure*}

For the energy value $\epsilon_-^0$ 
on the negative sheet we calculate the Hessian as
\begin{equation}
\Theta_- (0, 0) = \left[
\begin{array}{cc}
2 (\Omega - \gamma_x^2 / \omega_z )  	& 0				 \\
0								& 2 \Omega
\end{array}
\right].
\label{hessian.min}
\end{equation}
The determinant is given as $\det{\Theta_- (0, 0)} = 4 \Omega (\omega_z \Omega - \gamma_x^2 / \omega_z ) / \omega_z $, which is positive for $\omega_z > \gamma_x^2 / \Omega$.  Both diagonal entries in Eq. (\ref{hessian.min}) are positive for this case, meaning the Hessian is positive definite and the point $\epsilon_-^0$ is indeed a minimum. Visual inspection of the energy surface plots in Fig. \ref{fig:energy_surface_plots}(a, b) reveals this energy to be the unique global minimum in the symmetric phase.


For $\omega_z < \gamma_x^2 / \Omega$ the determinant  $\det{\Theta_- (0, 0)}$ becomes negative, giving an indefinite Hessian so that $\epsilon_-^0$ is now a saddle point as can be seen in Figs. \ref{fig:energy_surface_plots}(c, d).  This point goes from a stable critical point in the symmetric phase to an unstable critical point in the broken phase.
We also see in these figures that in this phase a new, twice degenerate ground state forms at the points $\pm \alpha_-$ as we discussed above.  These are the stable critical points in the broken phase.  In parallel with this transition, on the upper sheet the point $\epsilon_+^{\infty}$ goes from a global maximum to a saddle point at infinity in the broken phase.

We can obtain the explicit energy values associated with the broken phase global max/min points by plugging Eq. (\ref{crit.alpha.4points}) into Eq. (\ref{clas.energy.surface}) to find
\begin{equation}
\epsilon_{\pm}^{BP}
	   = \pm  \frac{1}{2 \gamma_x} \sqrt{(\omega_z^2 + \gamma_x^2)(\Omega^2 + \gamma_x^2)}.	   
\label{crit.energy.broken}
\end{equation}
The value on the lower sheet $\epsilon_-^{BP}$ gives the energy of the twice degenerate ground state.

\begin{figure*}
\includegraphics[width=0.90\textwidth]{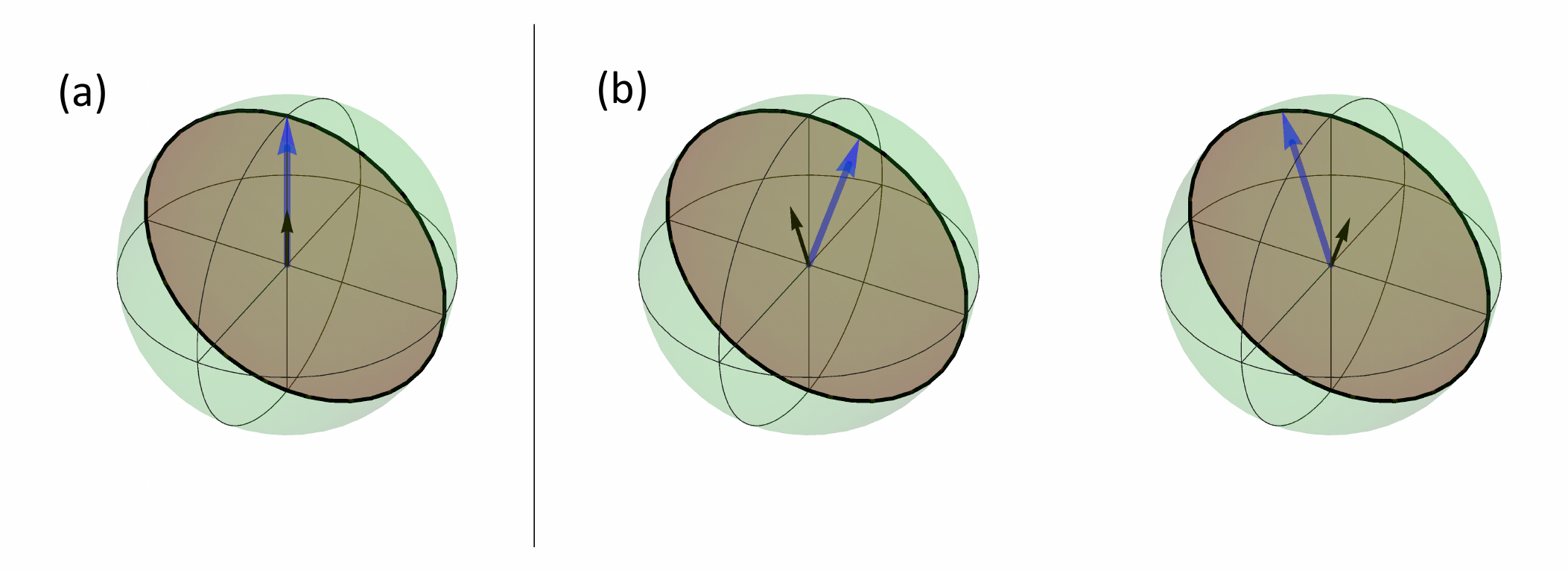}
\caption{(Color online) Qualitative descriptive diagrams for the ground state in (a) the symmetric phase, and (b) the broken phase.  The environment spin (large blue arrow) and central spin (black arrow)
reside in the $x$-$z$ plane (dark shading) in either case.  In the symmetric phase the arrows are aligned along the $z$ axis with the coupling effectively off.  In the broken phase, two degenerate ground states form with the components of the spin along the $x$-axis anti-aligned (the anti-alignment is only approximate for the case $\omega_z \neq \Omega$).}
\label{fig:sym.broken.spheres}
\end{figure*}

Plugging the critical values for $\alpha, \bar{\alpha}$ into Eqs. (\ref{xComp}-\ref{zComp}) we find the MFT values for the background spin in the ground state in either phase.  For the symmetric case with $\alpha = \bar{\alpha} = 0$ we find $\bra S_x \ket = \bra S_y \ket = 0$ while $\bra S_z \ket = N/2 = S$.  Meanwhile we can also find the 2 component eigenspinor associated with the symmetric phase ground state eigenvalue $\epsilon_-^0 = -(\Omega + \omega_z)/2$ in order to calculate $\bra \sigma_x \ket = \bra \sigma_y \ket = 0$ and $\bra \sigma_z \ket = 1$.  Hence we see that in the symmetric phase ground state, the central-spin and the background spin are both aligned along the positive $z$-axis, as represented in the qualitative diagram in Fig. \ref{fig:sym.broken.spheres}(a).

Meanwhile for the degenerate ground states in the broken phase we plug Eq. (\ref{crit.alpha.4points}) into Eq. (\ref{xComp}) to find
\begin{equation}
\frac{\bra S_x \ket}{ S } 
	= \pm \sqrt{ \frac{1 - \omega_z^2 \Omega^2 / \gamma_x^4 }{1 + \Omega^2 / \gamma_x^2} }
\label{mean.value.bp.x}
\end{equation}
for the $x$-component of the background spin as reported in \cite{Mermin,Lev2000,Lev&Muth2001}.
We see that the two degenerate ground states are associated with a left-right symmetry in terms of this component of the spin.
As this component was zero in the symmetric phase we can clearly interpret this as a second order quantum phase transition in which $S_x$ plays the role of the order parameter.  Indeed, we immediately read off the critical exponent for the order parameter from Eq. (\ref{mean.value.bp.x}) as $\beta = 1/2$, in agreement with the well-known MFT result \cite{Huang}.  Here $\beta$ describes the power-law scaling behavior of the order parameter in the vicinity of the QPT.

In a similar manner we may obtain
\begin{equation}
\frac{\bra S_y \ket}{ S } = 0, \;\;\;\;\; \;\;\;  
\frac{\bra S_z \ket}{ S } 
	= \sqrt{ \frac{1 + \omega_z^2 / \gamma_x^2 }{1 + \gamma_x^2 / \Omega^2} }.
\label{mean.values.bp}
\end{equation}
for the $y$ and $z$ components for the degenerate ground state.  We note from comparison of Eq. (\ref{crit.energy.broken}) and Eq. (\ref{mean.values.bp}) that the ground state energy in the broken phase is proportional to the mean value of the $z$ component of the background spin, according to 
$\epsilon^{\textrm{BP}}_- = - (\omega_z^2 + \gamma_x^2) \bra S_z \ket / 2 \Omega S$.

Finally we may obtain the expectation values for the small spin as
\begin{equation}
\bra \sigma_x \ket
	= \mp \sqrt{ \frac{1 - \omega_z^2 \Omega^2 / \gamma_x^4}
		{1 + \omega_z^2 / \gamma_x^2}}  , \;\;\;\;\; \;\;\;  
\bra \sigma_y \ket  =  0 , \;\;\;\;\; \;\;\; 
\bra \sigma_z \ket
	= \sqrt{ \frac{1 + \Omega^2 / \gamma_x^2 }{1 + \gamma_x^2 / \omega_z^2} }.
\label{mean.values.bp.s}
\end{equation}
Comparing the expression for $\bra \sigma_x \ket$ with $\bra S_x \ket / S$ from 
Eq. (\ref{mean.value.bp.x}) we see that the central-spin and the background spin in the degenerate ground state are exactly anti-aligned along the $x$-axis for the case $\omega_z = \Omega$ with equal energies.  For this special case we have 
$\bra S_x \ket / S = - \bra \sigma_x \ket = \pm \sqrt{\gamma_x^2 - \Omega^2} / \gamma_x$ (anti-alignment) and 
$\bra S_z \ket / S = \bra \sigma_z \ket = \Omega / \gamma_x$ (alignment).  For the case 
$\omega_z \neq \Omega$ the exact anti-alignment along $x$ and alignment along $z$ is broken, but it approximately holds for the case $\omega_z \sim \Omega$, which is our primary consideration here (we consider that all numerics presented in this paper fall into this parameter range).

The descriptive diagrams in Figs. \ref{fig:sym.broken.spheres}(a, b) may aid us in a qualitative understanding of the behavior of the ground state through this transition.
Fig. \ref{fig:sym.broken.spheres}(a) qualitatively represents the ground state in the symmetric phase with the central spin (black arrow) and environment spin (blue arrow) aligned in the positive $z$ axis.  Neither spin has an $x$-axis component, which is the coupling axis in our Hamiltonian Eq. (\ref{ham.mermin}).  

As we increase the relative strength of the coupling $\gamma_x$ and cross into the broken phase
we see in Fig. \ref{fig:sym.broken.spheres}(b) that the two spins each acquire a component that lies along the $x$ direction
(this diagram is drawn as a representation of the exact case $\omega_z = \Omega$).
For example in the left sphere the environment spin has acquired a component along the positive $x$-axis (consistent with the $+$ sign choice in Eq. (\ref{mean.value.bp.x})) while
the central spin has acquired a component along the negative $x$-axis (consistent with the $-$ sign choice in Eq. (\ref{mean.values.bp.s})).
The spins are anti-aligned in the $x$ direction as a result of the anti-ferromagnetic coupling ($\gamma_x > 0$) in the Hamiltonian (\ref{ham.mermin}) being `switched on' in the broken phase ground state.  Of course, there is also the equal-energy configuration with the central spin having a component along the positive $x$-axis and the environment spin pointing in the negative $x$-axis as depicted by the sphere on the right in Fig. \ref{fig:sym.broken.spheres}(b), which is the source of the broken phase degeneracy.

\begin{figure}
 \includegraphics[width=0.4\textwidth]{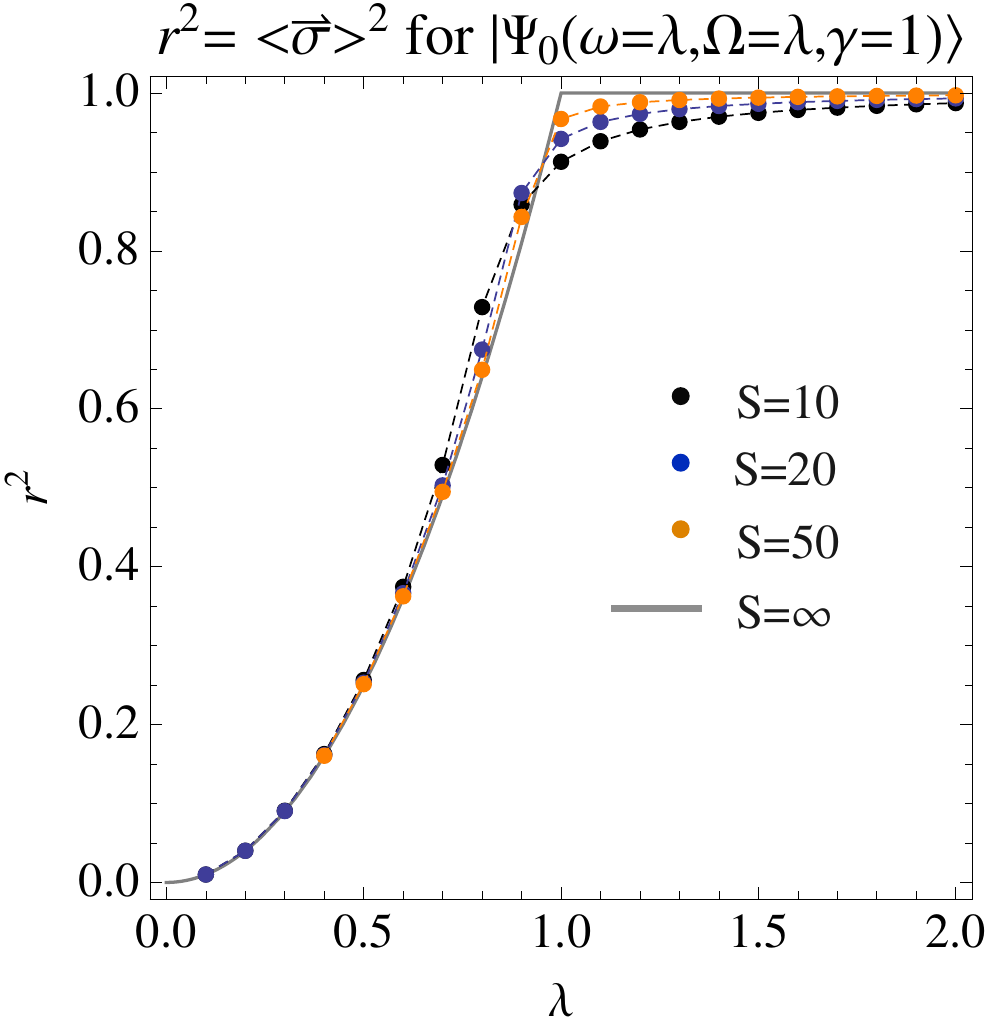}
 \caption{
Behavior of $r^2$ in the thermodynamic limit as a function of $\lambda \equiv \Omega = \omega_z$ in the case $\gamma_x = 1$ with numerics provided for the $S=10, 20$ and $50$ cases.
 }
\label{fig:entanglement}
\end{figure}

Finally, we are in a position to calculate the spin polarization $r^2$, the summed square of the components of the spin matrices expectation values (note that for the environment spin, we normalize this quantity by dividing by $S$).  This quantity is also known as the one-tangle.  Let us describe its properties using the central spin as an example: in the case that the central spin is in a pure state, we shall have $r^2 = 1$.  In this case, $r^2$ is equivalent to the usual Bloch sphere radius.  However, to extend the definition of this quantity to the mixed state case, one must allow for $r^2 < 1$, in which situation we can uniquely describe the system properties in terms of the \emph{Bloch Ball}
\footnote{
The state of a two level system can be unambiguously represented by
a point in the interior of or on the surface of the unit sphere. This sphere is the boundary
of a unit ball, the Bloch Ball. A point on the interior of this ball represents
unambiguously the matrix density of a mixed state. When a two level
system is entangled with the bath, a mixed state is obtained upon tracing
the bath degrees of freedom. Therefore the representation of the
central spin within the Bloch ball gives a measure of its entanglement
with the bath
\cite{Mosseri&Dand}.}.
Then $r^2 \le 1$ serves as a precise measure of the entanglement present in the system; the smaller the value of $r^2$ the greater is the entanglement in the system, with $r = 0$ yielding the completely mixed state  \cite{Mosseri&Dand}.





For both the central spin and the environment spin, $r^2$ sums to one in the symmetric phase so there is no entanglement here.
However, for the degenerate phase ground state (which is a superposition of the two states with eigenvalue $\epsilon_-^{\textrm{BP}}$) the $x$-component contributions will cancel, such that $r^2$ is given simply by the square of the $z$-component of the spin matrix expectation value on both sides of the QPT.  This quantity is less than one throughout the broken phase, but approaches one at the QPT as can be seen immediately from Eqs. (\ref{mean.values.bp}) and (\ref{mean.values.bp.s}).
As an illustration of this, we plot $r^2$ for the central-spin in Fig. \ref{fig:entanglement} as a function of 
$\lambda \equiv \Omega = \omega_z$.  When $\lambda$ crosses $\gamma_x = 1$ in this diagram, we cross the QPT.  From the numerics, we see that as the number of background spins is increased, the $r^2$ value approaches the ideal behavior in the thermodynamic limit.


\subsection{Transition between overlapping and non-overlapping energy surfaces}\label{SEC:energy.surface.MFT.overlap}

For a certain range of parameter space the two energy surfaces may overlap along the energy axis, as we have already seen in Figs. \ref{fig:energy_surface_plots}(b, d).
This also leads to a type of degeneracy in the system as different states on different sheets have equal energy.

Note that $\epsilon_-^\infty$ forms the global maximum on the lower sheet and $\epsilon_+^0$ forms the global minimum on the upper sheet (this statement is true on both sides of the QPT).
Hence at the point in parameter space where the transition to overlapping sheets occurs, these two points represent the first overlapping energy values. 
Therefore, making use of Eqs. (\ref{crit.energy.alpha.zero}) to set $\epsilon_+^0 = \epsilon_-^\infty$ immediately yields
$\omega_z = \Omega$ as the transition point.  Notice that these points must meet on the origin of the energy axis.  

\begin{figure}
 \includegraphics[width=0.8\textwidth]{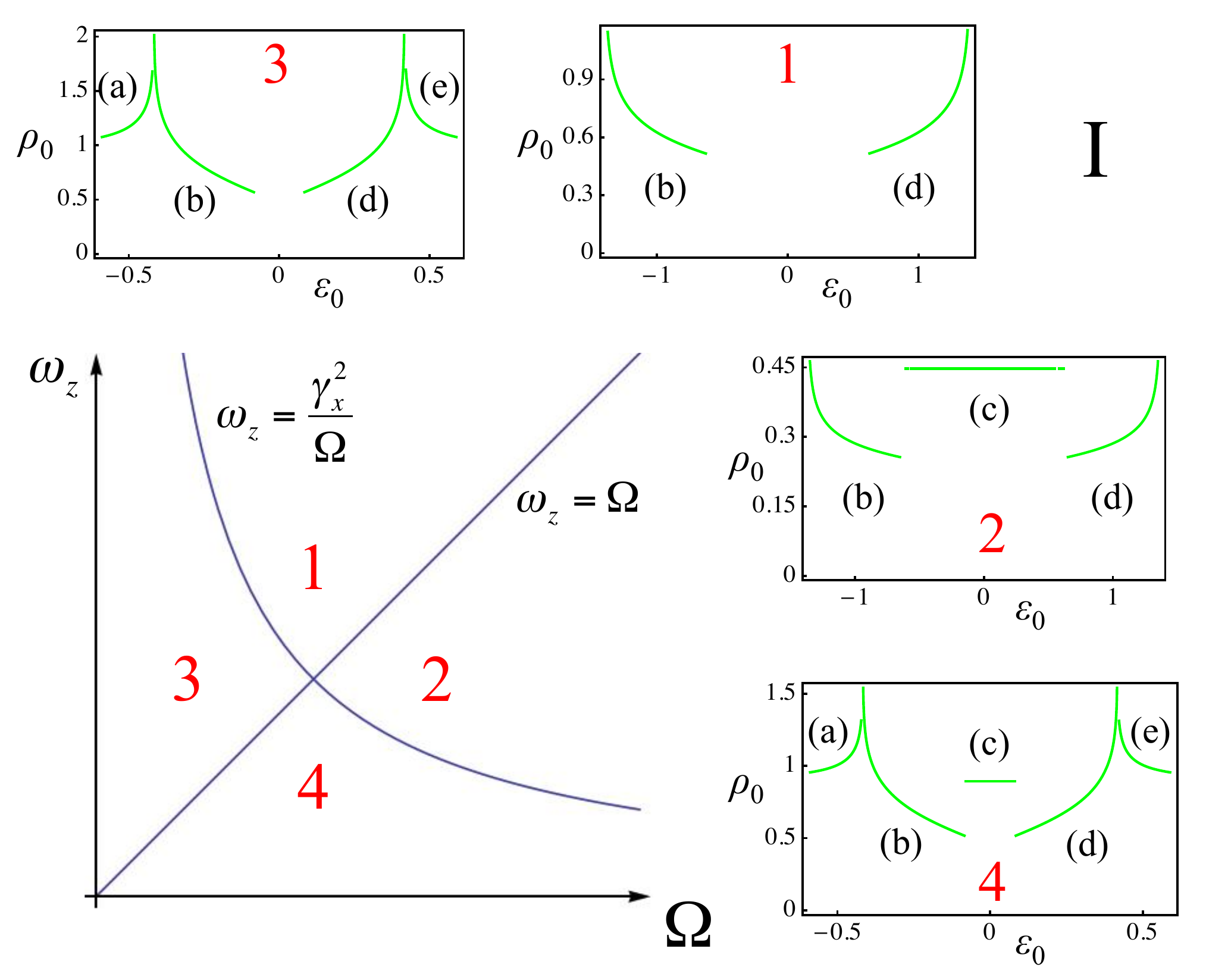}
 \caption{(Color online) Here we show the phase diagram for Quadrant I from Fig. \ref{fig:phase.quads} in greater detail.  We show the curve $\omega_z = \frac{\gamma_x^2}{\Omega}$ that denotes the quantum phase transition, as well as the line $\omega_z = \Omega$ which further divides the phase diagram into the odd-numbered regions (1, 3) with non-overlapping energy sheets and the even-numbered regions (2, 4) with overlapping sheets.  We have chosen the value $\gamma_x = 1$.  In the four surrounding panels we plot the density of states $\rho_0$ against $\epsilon_0$ for representative values in each of the four regions.  We choose the values as: Region 1 ($\Omega = 3/4, \omega_z = 2$); Region 2 ($\Omega = 2, \omega_z = 7/10$); Region 3 ($\Omega = 1/3, \omega_z = 1/2$); Region 4 ($\Omega = 1/2, \omega_z = 1/3$), with $\gamma_x = 1$ in each case.}
 \label{fig:phase.regions}
 \end{figure}

Having obtained the condition $\omega_z = \Omega$ as the point where the transition to overlapping energy sheets occurs, we now divide our phase diagram for quadrant I into four distinct regions, as
in Fig. \ref{fig:phase.regions}.  Here we have Regions 1 and 2 as the broken phase region and 3 and 4 as the symmetric phase.  The curve ($\omega_z = \gamma_x^2 / \Omega$) acts as the dividing line between these regions.  These two portions of the phase diagram are then further subdivided according to Regions 1 and 3 as those without overlap  and Regions 2 and 4 as those with overlapping energy sheets, with the line $\omega_z = \Omega$ acting as the dividing line in the latter case.

\section{Beyond Mean Field Theory: Spectra in the Majorana Representation}\label{SEC:diffeq.thermo}

In this section we will relax the MFT approximation from the previous section and explicitly write the form of the Schr\"odinger equations in the Majorana representation as two de-coupled second-order differential equations.  We will present numerical solutions to these equations on the Majorana sphere.
We will then introduce the  function $G^{s,a}(\alpha)$, as the logarithmic derivative of the Majorana polynomials $\phi_{s,a}(\alpha)$.  The roots of the Majorana polynomials will be transferred to the poles of the function, $G^{s,a}(\alpha)$, while the second-order equations will be transformed to first-order equations.  We will then obtain an expansion for $G^{s,a}(\alpha)$ that allows us to obtain the integrated density of states in the thermodynamic limit in Sec. \ref{SEC:anal.dos}.

\subsection{Schr\"odiner equation in terms of Majorana polynomials}\label{SEC:diffeq.thermo.maj}

Having obtained the matrix form of the Hamiltonian (\ref{eigen.eqns.matrix.sa}) in the symmetrized basis, we can immediately re-write the Schr\"odinger equation in terms of the Majorana polynomials
\begin{equation}
H {\bf \phi} (\alpha) = \epsilon {\bf \phi} (\alpha)
\label{sym.sch.eqn}
\end{equation}
in the form of two coupled first-order equations $A_- \phi_s = (\omega_z / 2) \phi_a$ and $A_+ \phi_a = (\omega_z / 2) \phi_s$, in which
\begin{equation}
A_\pm \equiv \left(
\epsilon -  \Omega { S_z \over 2S} \pm \gamma_x { S_x \over 2S}
\right).
\label{A_pm.defn}
\end{equation}
We can immediately de-couple these equations to obtain the second-order differential equations
\begin{eqnarray}
A_+ A_- \phi_s = {\omega_z^2 \over 4} \phi_s, \nonumber \\
A_- A_+ \phi_a = {\omega_z^2 \over 4} \phi_a
\label{decoupled.sch.eqn}
\end{eqnarray}
in symbolic form.   Now, making use of the expressions (\ref{S.ops.coherent.rep}) for the $S$ operators in the coherent states representation and the definitions of $A_\pm$ given in Eq. (\ref{A_pm.defn})
 we can obtain the explicit form of these second order equations as
\begin{equation}
 \left[ \frac{P_2^{s,a} (\alpha) }{(2S)^2} \partial_\alpha^2 + \frac{P_1^{s,a} (\alpha) }{2S} \partial_\alpha + P_0^{s,a} (\alpha)
 \right] \phi_{s,a} = 0
\label{2nd.diffeq.sym.anti}
\end{equation}
with the coefficients at each order given by
\begin{eqnarray}
 P_0^{s,a} & = & \frac{\left((1-2 S) \alpha ^2-1\right) \gamma _x^2 
 \pm 2 \alpha  \Omega  \gamma _x
 +2 S (2 \epsilon -\omega +\Omega ) (2 \epsilon +\omega +\Omega ) }{8 S} \nonumber \\
 P_1^{s,a} & = & \frac{(2 S-1) \alpha  \left(\alpha ^2-1\right) \gamma
   _x^2 \mp \left(\alpha ^2+1\right) \Omega  \gamma _x+2 \alpha 
   \Omega  (\Omega -2 S (2 \epsilon +\Omega ))}{4 S}  \nonumber \\
 P_2^{s,a} & = & \frac{1}{4} \left(4 \alpha ^2 \Omega ^2-\left(\alpha
   ^2-1\right)^2 \gamma _x^2\right).
\label{P012.sa}
\end{eqnarray}
Note that the symmetric and anti-symmetric expressions differ only with respect to terms that have no 
$S$ dependence.  Hence, in the thermodynamic limit these terms drop out and the equations for the two sectors will precisely agree.  This implies that in the thermodynamic limit $\phi_s$ and $\phi_a$ have the same functional dependence (hence also that $\psi_+$ and $\psi_-$ must have the same form up to a constant in this limit).  Therefore, we introduce the notation $\phi^0 \equiv \lim_{S \rightarrow \infty} \phi^s = \lim_{S \rightarrow \infty} \phi^a $.  In Sec. (\ref{SEC:diffeq.thermo.Gfcn}) below we will explicitly consider this limit, after we have introduced the $G$ function.

\subsection{Numerical plots of solutions on the Majorana sphere}\label{SEC:diffeq.thermo.Mspheres}

Here we plot some numerical results 
to illustrate our method and to gain insight into the physics of the system.  This will also facilitate our discussion of the integrated density of states calculation in Sec. \ref{SEC:anal.dos} and our analysis of the expectation value of system observables in Sec. \ref{SEC:observables}.
To conduct our simulations for a given set of parameter values, first we find the form of our (two) Majorana polynomials (\ref{psi.coherent}) by numerically solving the Mermin model Hamiltonian (\ref{ham.mermin}).  Then we project the roots of the polynomials onto the Majorana sphere using an inverse stereographic projection of the form
\begin{equation}
\pi (\alpha) = \frac{1}{1 + \bar{\alpha} \alpha} 
\left(
\begin{array}{c}
\alpha + \bar{\alpha}  \\
(\alpha - \bar{\alpha}) / i \\
\alpha \bar{\alpha} - 1
\end{array}
\right).
\label{sphere.projection}
\end{equation}

\begin{figure*}
\includegraphics[width=0.90\textwidth]{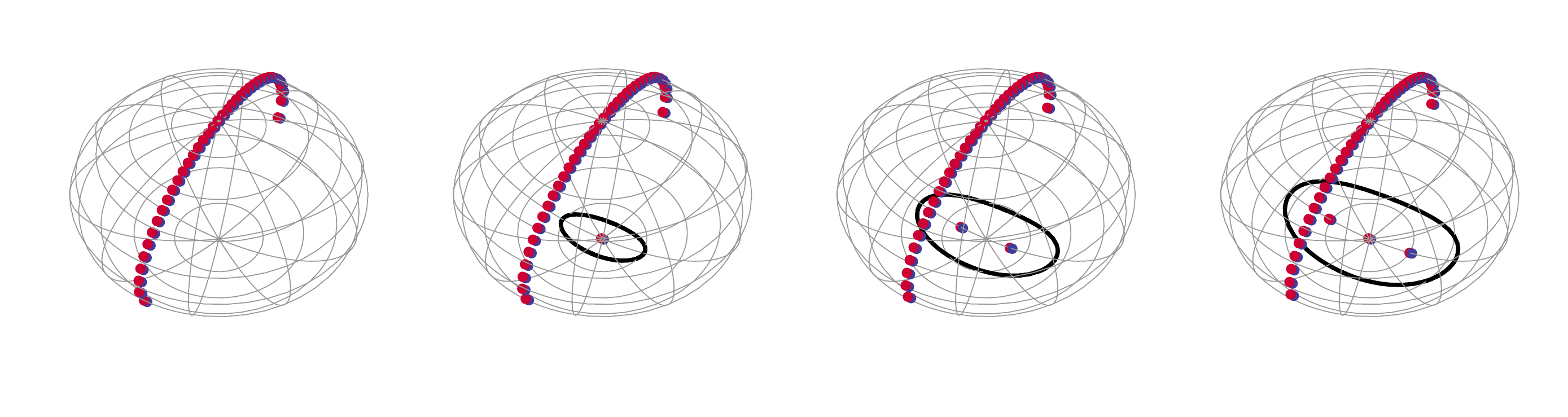}
\caption{(Color online) Projection of the roots of the Majorana polynomial for low-excitation states in Region 1.
The projections of the solutions to the $\phi_s (\alpha)$ sector are shown in blue and those of the $\phi_a (\alpha)$ sector are shown in red.
Here we have chosen the numerical values $\omega_z = 2.5$, $\Omega = 0.5$ and $\gamma_x=1$ (Region 1) with the number of background spins $N=40$.  From left to right, we have the $k=1$ state (ground state), followed by $k=2, 3, 4$.  The poles on the lower hemisphere line up along a straight-line contour on the real axis as we move from left to right.  In each case, this contour is surrounded by the projected level energy surface, shown as the black circular curve.
}
\label{fig:spheres.reg1.gs}
\end{figure*}

\begin{figure*}
\includegraphics[width=0.90\textwidth]{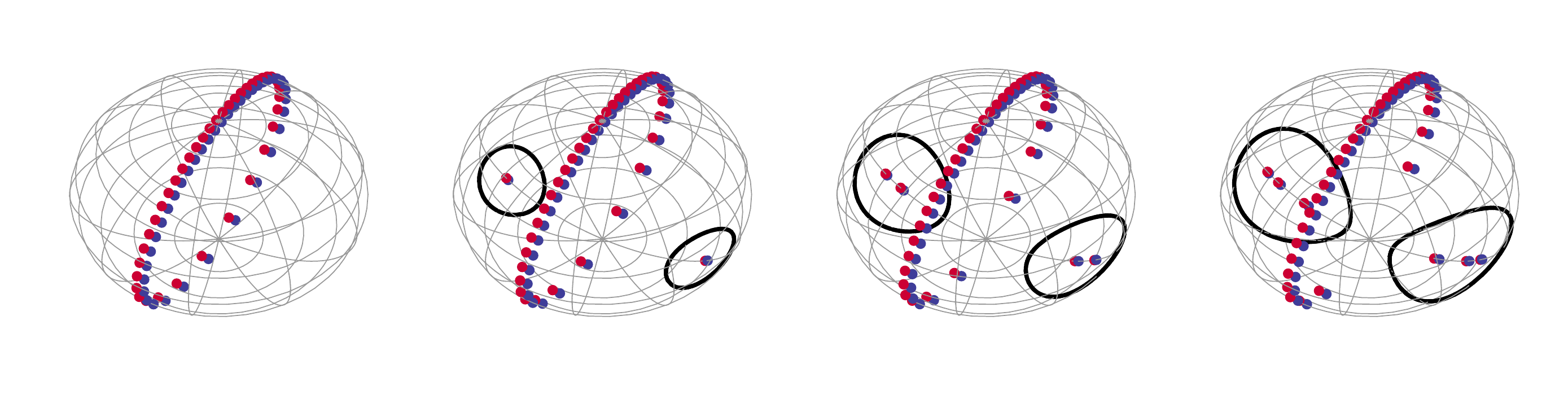}
\caption{(Color online) Projection of the roots of the Majorana polynomial for low-excitation states in Region 3.
The projections of the solutions to the $\phi_s (\alpha)$ sector are shown in blue and those of the $\phi_a (\alpha)$ sector are shown in red.
Here we have chosen the numerical values $\omega_z = 0.95$, $\Omega = 0.5$ and $\gamma_x=1$ (region 3) with the number of background spins $N=40$.  From left to right, we have the $k=1$ state (ground state), followed by $k=3, 5, 7$ (we present odd numbers here to avoid a minor technicality).  The two distinct level energy surfaces are shown in the $k=3, 5, 7$ spheres as the black curves.
}
\label{fig:spheres.reg3.gs}
\end{figure*}

\begin{figure*}
\includegraphics[width=0.90\textwidth]{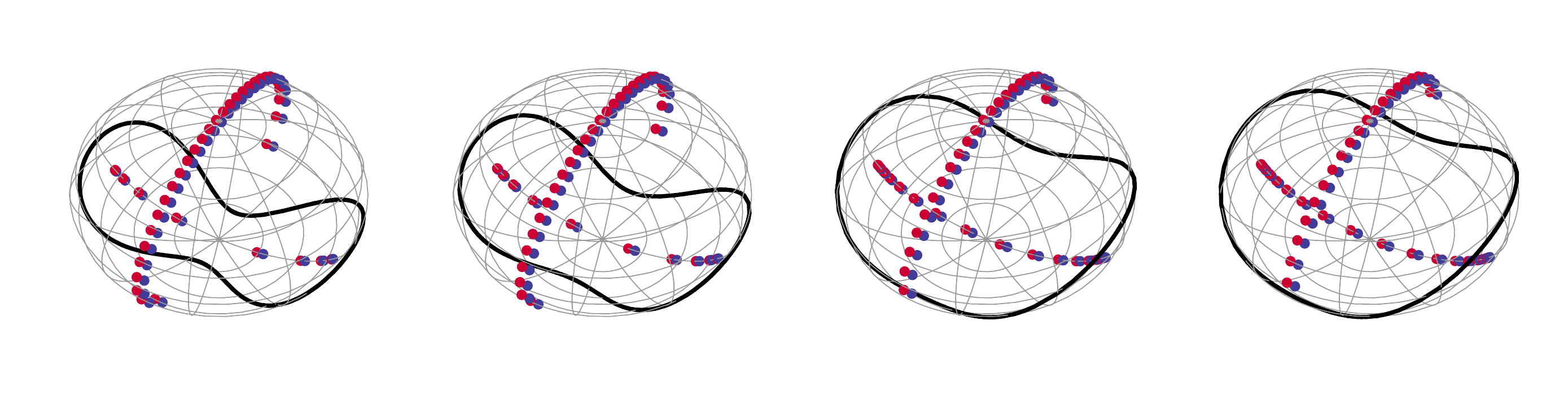}
\caption{(Color online) Projection of the roots of the Majorana polynomial for excitation states near the saddle point in Region 3.
The projections of the solutions to the $\phi_s (\alpha)$ sector are shown in blue and those of the $\phi_a (\alpha)$ sector are shown in red.
Here we have chosen the numerical values $\omega_z = 0.95$, $\Omega = 0.5$ and $\gamma_x=1$ (region 3) with the number of background spins $N=40$.  From left to right, we have the $k=9, 11, 17, 19$ states.  
The level energy surface is shown in each sphere as the black curve.
}
\label{fig:spheres.reg3.qpt}
\end{figure*}

\begin{figure*}
\includegraphics[width=0.90\textwidth]{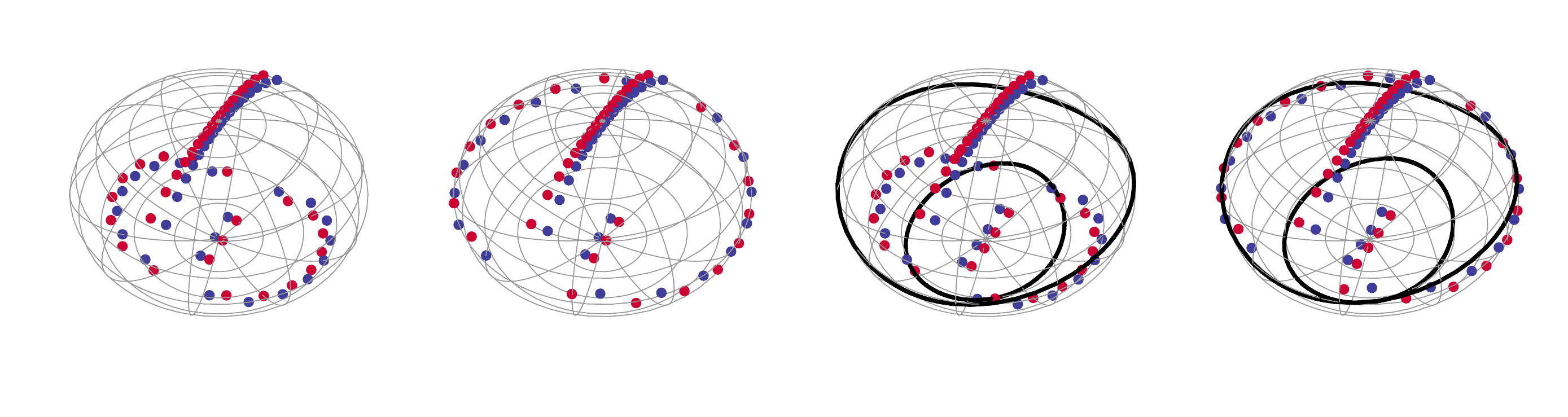}
\caption{(Color online) Projection of the roots of the Majorana polynomial for excitation states in the portion of the spectrum with overlapping sheets in Region 2.
The projections of the solutions to the $\phi_s (\alpha)$ sector are shown in blue and those of the $\phi_a (\alpha)$ sector are shown in red.
Here we have chosen the numerical values $\omega_z = 7/10$, $\Omega = 2$ and $\gamma_x=1$ (region 2) with the number of background spins $N=40$.  From left to right, we have the $k=24, 25, 26, 27$ states.  Notice the alternating behavior of the ``halo'' shape, in the first sphere appearing on the lower hemisphere, then on the upper hemisphere in the second, and so forth.}
\label{fig:spheres.reg2.overlap}
\end{figure*}

One interesting feature of the Majorana sphere is that the arrangement of the roots projected onto the sphere reflects certain elements of the classical energy surface.  In the present case, we will see parallels between the Majorana sphere projection and the classical energy surface diagrams presented in Fig. \ref{fig:energy_surface_plots}.  To illustrate this point, in certain cases below we also project the level energy surfaces (the black curves in Fig. \ref{fig:energy_surface_plots}) onto the sphere.  These are also visible as black curves in Figs. \ref{fig:spheres.reg1.gs} - \ref{fig:spheres.reg2.overlap}.

In Fig. \ref{fig:spheres.reg1.gs} we show projections of the numerically obtained roots for the low-excitation states near the ground state for values representative of Region 1.  
We use here the values $\omega_z = 2.5$, $\Omega = 0.5$ and $\gamma_x=1$  with the number of background spins $N=40$.
The roots of the $\phi_s (\alpha)$ polynomial are shown in blue and those of the $\phi_a (\alpha)$ polynomial are shown in red.  The first sphere on the far left represents the ground state, while the next three represent the first three excited states, respectively.

In the case of the ground state at energy $\epsilon_-^0$ note that all of the poles line up along the imaginary axis in the upper hemisphere (in the complex plane, these poles extend to infinity along the positive and negative imaginary axis).  As we move rightward we see that at each consecutive excited state a pair of poles vanishes from the upper hemisphere and appears at or in the vicinity of the origin (lower hemisphere).  As they accumulate, we see that these poles are lining up on the real axis on the lower hemisphere as we move to higher excitation states.  This process continues until all of the poles appear on the real axis, which occurs when we reach the upper edge of the lower energy sheet.

This is useful for us because we can use the number of poles on a given (straight-line) curve as a way to index the polynomial solutions.  Further on, we will define a complex integration contour around the poles on the real axis that allows us to `count' which state we are in, and therefore we can calculate the integrated density of states as a function of energy $\epsilon$.  As we consider other phase regions or different cases (such as overlapping sheets, etc.) we will find that this contour where the poles line up changes, however, some such contour(s) exists (exist) for all parameter values.  It is this fact that allows us to obtain an analytic expression for the integrated density of states in each case.

For the present case of Fig. \ref{fig:spheres.reg1.gs}, note further that the poles appearing in the lower hemisphere are enveloped by the level energy surface (black curve in spheres 2, 3 and 4).  Here it is useful to think of moving rightwards among the spheres in Fig. \ref{fig:spheres.reg1.gs} as moving upwards on the classical energy surface 
in Fig. \ref{fig:energy_surface_plots}(a).  Hence we may think of the poles on the lower hemisphere as `nodes' that are expanding to fill the level surfaces, which themselves expand with increasing energy.

In Fig. \ref{fig:spheres.reg3.gs} we have given projections on the Majorana sphere for low-excitation states in Region 3 for the critical regime of the lower energy sheet.
From left to right in this case we have the (broken phase) ground state at $\epsilon_-^{\textrm{BP}}$, then the third, fifth and seventh excited states, respectively.
In this case, the poles line up on two distinct contours along the real axis, but separated from the origin.
Indeed these two contours are associated with the two potential wells on the lower energy sheet that we encountered previously in Fig. \ref{fig:energy_surface_plots}(c).  These potential wells are represented in Fig. \ref{fig:spheres.reg3.gs} by the two circular projected level energy surfaces enveloping the poles along the two straight-line contours.  

In Fig. \ref{fig:spheres.reg3.qpt} we present the Majorana sphere projections also for Region 3, but in this case we are passing the spectrum saddle-point at $\epsilon_-^0$ ( see Fig. \ref{fig:energy_surface_plots}(c) ).  
Here it is clear that the two separate contours considered in Fig. \ref{fig:spheres.reg3.gs}
merge above the saddle-point to form a single contour. 
For the two spheres on the left of Fig. \ref{fig:spheres.reg3.qpt} we see that the two circular level surfaces have come together to form the united `figure 8' shape from Fig. \ref{fig:energy_surface_plots}(c).


Finally in Fig. \ref{fig:spheres.reg2.overlap} we plot four spheres as an example of the behavior in the overlapping portion of the spectrum for Region 2
(here we have chosen to illustrate the $k=24, 25, 26, 27$ excitation states).  We have repressed the level energy surface curves in the two leftmost spheres.
In this case, we see that as we move to higher excitation states, the ``halo'' shape appearing on the lower hemisphere in the leftmost sphere will appear alternately on the upper hemisphere (second sphere from the left) then return to the lower hemisphere (third sphere from the left) and so on.

Again we can make sense of this behavior in terms of the level energy surfaces.  Returning to the overlapping portion of the diagram in Fig. \ref{fig:energy_surface_plots}(b) we see that the level energy surfaces appear as two circles: one small, inner circle (on the upper sheet) and one larger, outer circle (on the lower sheet).  These represent two surfaces at the same energy that are not related by symmetry, similar to the previous studies of the LGM model (see the lower two spheres $(e^-, e^+)$ in Fig. 4 of \cite{Ribeiro2}).

These two surfaces are projected onto the two rightmost spheres in Fig. \ref{fig:spheres.reg2.overlap}.
Here we see that the halo shape seems to be alternating between the two sheets, lying more closely to the level surface in the lower hemisphere (upper sheet) in the third sphere from the left, while lying more closely to the level surface in the upper hemisphere (lower sheet) in the rightmost sphere.
We will see that this alternating behavior again occurs in terms of the system observables in Sec. \ref{SEC:observables}, in particular see Figs. \ref{fig:exp.vals.R2} and \ref{fig:exp.vals.R4} .

\subsection{$G$ function and expansion in terms of particle number}\label{SEC:diffeq.thermo.Gfcn}

Let us now introduce the logarithmic derivative functions of our symmetric and anti-symmetric sector wave functions as
\begin{equation}
G^{s,a} = \frac{1}{2S} \partial_\alpha \log{\phi^{s,a}} = \frac{1}{2S} \sum_{k=1}^d \frac{1}{\alpha - \alpha_k^{(s,a)} }.
\label{G.sym.anti}
\end{equation}
where $\alpha_k^{(s,a)}$ are respectively the roots of the symmetric and anti-symmetric polynomials.
According to this transformation, the roots of the Majorana polynomials have been transferred to poles of the functions $G^{s,a}(\alpha)$.
While $G^{s,a}(\alpha)$ is not exactly a Green's function, its roots serve to keep track of which state the system is in; hence, we can use it to calculate the integrated density of states.

Introducing the $G^{s,a}(\alpha)$ functions allows us to re-write the second order differential equations (\ref{2nd.diffeq.sym.anti}) as the first order equations
\begin{equation}
 P_2^{s,a} (\alpha) \left[ (G^{s,a} (\alpha))^2 + \frac{1}{2S} \partial_\alpha G^{s,a} (\alpha) \right] 
 	+ P_1^{s,a} (\alpha) G^{s,a} (\alpha) + P_0^{s,a} (\alpha)   =  0.
\label{1st.diffeq.sym.anti}
\end{equation}
Then, in order to take the thermodynamic limit, we can expand $G^{s,a}$ and $\epsilon$ in the form of
\begin{equation}
 G^{s,a} = \sum_{i=1}^\infty \frac{G_i^{s,a}}{S^i},
 \epsilon = \sum_{i=1}^\infty \frac{\epsilon_i}{S^i}.
\label{G.expansion.sa}
\end{equation}
With these expansions equation (\ref{G.sym.anti}) reduces to a quadratic equation for $G_0$ at the thermodynamic limit $S \rightarrow \infty$.  These equations take the form
\begin{equation}
 P_2^{\infty} (\alpha) (G_0 (\alpha))^2
 	+ P_1^{\infty} (\alpha) G_0 (\alpha) + P_0^{\infty} (\alpha)   =  0.
\label{G0.quadratic.sa}
\end{equation}
with the coefficients here obtained from Eqs. (\ref{P012.sa}) as
\begin{eqnarray}
 P_0^{\infty} (\alpha) = \frac{1}{4} \left(4 \epsilon ^2+4 \Omega  \epsilon -\omega
   ^2+\Omega ^2-\alpha ^2 \gamma _x^2\right)   \\
 P_1^{\infty} (\alpha) = \frac{1}{2} \alpha  \left(\left(\alpha ^2-1\right) \gamma _x^2-2
   \Omega  (2 \epsilon +\Omega )\right)   \\
 P_2^{\infty} (\alpha) = P_2^{s,a} (\alpha)
 	= - \frac{ \gamma_x^2}{4}  ( \alpha^2 - \beta_+^2 )( \alpha^2 - \beta_-^2 )
\label{P012.thermo}
\end{eqnarray}
where
\begin{equation}
\beta_{\pm} = \frac{\Omega \pm  \sqrt{\Omega^2 + \gamma_x^2} }
	{\gamma_x}.
\label{beta.12}
\end{equation}
We emphasize that while we are still working in the thermodynamic limit here, we have relaxed the MFT assumption from Sec. \ref{SEC:energy.surface}, hence we are working at a first-order level in the immediate approximation scheme.

\subsection{Solution for $G_0$ at the thermodynamic limit}\label{diffeq.thermo.G0}

We can explicitly write the solution of the quadratic equation (\ref{G0.quadratic.sa})  for $G_0$ as
\begin{equation}
 G_0^\pm (\alpha ; \epsilon_0)
	= \frac{\alpha \left[2\Omega (2\epsilon_0 - \Omega) + (1 - \alpha^2)\gamma_x^2 \right] \pm 2 \sqrt{Q(\alpha; \epsilon_0)}}
	{4 P_2^{\infty}(\alpha)}
\label{G0}
\end{equation}
in which $Q(\alpha ; \epsilon)$ is itself a quadratic expression in $\alpha^2$ given by
\begin{equation}
 Q(\alpha; \epsilon) = \Gamma_- (\epsilon) (\alpha^2 - r_+^2)(\alpha^2 - r_-^2)
\label{Q.roots}
\end{equation}
with
\begin{equation}
\Gamma_- (\epsilon) = 
(2 \epsilon - \omega_z - \Omega ) (2 \epsilon + \omega_z - \Omega )
   \gamma _x^2
\label{Q.Gamma_pm}
\end{equation}
The quadratic roots of $Q(\alpha ; \epsilon)$ are given by
\begin{equation}
r_\pm^2 = \frac{ \gamma_x (\epsilon^2 - \bar{\epsilon}^2) 
			\pm \omega_z \Omega \sqrt{(\epsilon_{-}^{BP})^2 - \epsilon^2} 
				  }
			{\gamma_x (\epsilon - \epsilon_+^\infty)(\epsilon - \epsilon_-^\infty)},
\label{alpha_pm.2}
\end{equation}
in which
\begin{equation}
\bar{\epsilon}^2 = \frac{(\omega_z^2 + \Omega^2) \gamma_x^2 + 2 \omega_z^2 \Omega^2 }
					{4 \gamma_x^2}
\label{epsilon.bar}
\end{equation}
(we repress the $\epsilon$ dependence in $r_\pm (\epsilon)$ to avoid too heavy of notation).  
Recall that $\epsilon_{\pm}^{\infty}$ and  $\epsilon_-^{BP}$ are given in 
Eqs. (\ref{crit.energy.alpha.infty}) and (\ref{crit.energy.broken}), respectively.
Since $Q(\alpha; \epsilon_0)$ is under the radical in Eq. (\ref{G0}), these roots describe two $\epsilon$-dependent branch cuts in the complex $\alpha$ plane; $\left[ -r_+, r_+ \right]$ and $\left[ -r_-, r_- \right]$.  These cuts are precisely the contours discussed in Sec. \ref{SEC:diffeq.thermo.Mspheres} along which the roots will organize themselves with increasing 
$\epsilon$ \footnote{Notice that in the portion of the spectrum correlating to the broken phase, the two contours will partially overlap; the non-overlapping portion correlates to the line on which the roots are gathering in the lower hemisphere in Fig. \ref{fig:spheres.reg3.gs}.}.


\section{Analytic calculations for the Density of States}\label{SEC:anal.dos}

We have now developed the theoretical formalism that will allow us to perform the calculation of the integrated density of states, where the density of states itself is given by
\begin{equation}
\rho(\epsilon) = \sum_n \delta(\epsilon-\epsilon_n)
\label{dos.defn}
\end{equation}
for the Mermin model in the thermodynamic limit, where $\epsilon_n$ are the system energy eigenvalues.  This calculation is the primary goal of the present work.

In order to facilitate our calculation, we define the density of poles $\mathcal I(\epsilon)$ along the contour $C_1$ that follows one of the branch cuts in the complex plane 
as
\begin{equation}
\mathcal I^{s,a} (\epsilon) 
	=  \frac{p}{2 S} 
	= \frac{1}{2 i \pi} \int_{\tilde{C}_1} G^{s,a} (\alpha; \epsilon) d \alpha
\label{pole.density}
\end{equation}
in which $p$ is the total number of poles along $C_1$ and $\tilde{C}_1$ is a contour that surrounds $C_1$.  $C_1$ itself follows one of the branch cuts in the complex plane, although this contour will change for different portions of the spectrum.

Essentially what we have done is that we have transferred the zeroes of the Majorana polynomials into the poles of the functions $G^{s,a}(\alpha; \epsilon)$.  For given parameter values $\Omega, \omega_z$ and $\gamma_x$, as we increase $\epsilon$ the positions of the poles will change as we saw in Sec. \ref{SEC:diffeq.thermo.Mspheres}.
For example, as we increase the energy from the ground state in Region 1, 
these poles will gradually accumulate on the real axis in the lower hemisphere as in 
Fig. \ref{fig:spheres.reg1.gs}.
Hence, by choosing our contour $C_1$ along the the branch cut on the real axis from $-r_+$ to $r_+$, the quantity in Eq. (\ref{pole.density}) measures the accumulation of poles with increasing $\epsilon$.  Since the pole density $\mathcal I^{s,a} (\epsilon)$ essentially counts the poles as we move from one state to the next, this quantity can be related to the integrated density of states 
\begin{equation}
\mathcal N (\epsilon) \equiv \int_{-\infty}^\epsilon  \rho(\epsilon^\prime) d \epsilon^\prime
\label{int.dos.defn}
\end{equation}
as in the following section.

\subsection{Explicit integral form of integrated density of states}\label{SEC:anal.dos.integral}

In the thermodynamic limit $S \rightarrow \infty$ it can be shown that the integrated density of states $\mathcal N (\epsilon)$ can be written as \cite{Ribeiro2}
\begin{equation}
\mathcal N_0 (\epsilon_0) = \lim_{S \rightarrow \infty} \mathcal N^{s,a}(\epsilon) 
	=  \lim_{S \rightarrow \infty} I^{s,a}(\epsilon) = \mathcal I_0 (\epsilon_0)
\label{equation}
\end{equation}
in which $\epsilon_0$ is the first-order term for the energy defined in Eq. (\ref{G.expansion.sa}) and
\begin{equation}
\mathcal I_0 (\epsilon_0) 
	= \frac{1}{2 \pi i} 
		\int_{C_1} d \alpha \left[ G_0^+ (\alpha; \epsilon_0) - G_0^- (\alpha; \epsilon_0) \right].
\label{dos.thermo}
\end{equation}
The analytic portion of the integration will cancel out, hence we have
\begin{equation}
\mathcal I_0 (\epsilon_0) 
	= \frac{1}{2 \pi i} \int_{C_1} d \alpha \frac{\sqrt{Q(\alpha; \epsilon_0)}}{P_2^\infty (\alpha)}
		= - \frac{2}{i \pi \gamma_x^2} 
		    \int_{C_1} d \alpha \frac{ \sqrt{\Gamma_- (\epsilon_0) (\alpha^2 - r_+^2)(\alpha^2 - r_-^2)} }
								{(\alpha^2 - \beta_1^2)(\alpha^2 - \beta_2^2)}.
\label{dos.Q}
\end{equation}
This is our standard integral form for the integrated density of states.  In each case below, we can transform this integral in terms of the complete elliptic integral of the first kind
\begin{equation}
K(k) 
	= \int_0^{1} \frac{dx}{\sqrt{(1 - x^2)(1 - k x^2)}} 
\label{K.k}
\end{equation}
and the complete elliptic integral of the third kind
\begin{equation}
\Pi (n, k)
	= \int_0^{1} \frac{dx}{(1 - n x^2 ) \sqrt{(1 - x^2)(1 - k x^2)}}.
\label{Pi.n.k}
\end{equation}

\subsection{Analytic form for integrated density of states by region}\label{sec:anal.dos.expr}

Below we present the results of the integration given in Eq. (\ref{dos.Q}) for each phase region in Fig. \ref{fig:phase.regions}.
We find that there are five different zones (a-e) corresponding to five different types of behaviors in the spectrum.
Not all of these zone types actually occur in each phase region; only Region 4 is comprised of all five.
These zones are shown in terms of the density of states for representative cases for all four regions in Fig. (\ref{fig:phase.regions}).
In the Appendix \ref{app:DOS.calc} we show how to obtain the results presented in zone (d) by an analytic transformation of the integrand given in Eq. (\ref{dos.Q}), which is representative of the general method.

\subsubsection{Region 1 (symmetric, non-overlapping sheets)}\label{sec:anal.dos.expr.reg1}

Let us first consider the case of Region 1 with non-degenerate ground state and non-overlapping energy surfaces.  This is given by $\omega_z > \gamma_x^2 / \Omega$ and $\omega_z > \Omega$ in Fig. \ref{fig:phase.regions}.  We find that the spectrum is comprised of zone (b) behavior on the lower sheet and zone (d) behavior on the upper sheet.

(b) $ - \frac{\omega_z + \Omega}{2} < \epsilon_0 < \frac{- \omega_z + \Omega}{2} $ (Lower sheet).
The integrated density of states in this zone can be written in one of two ways, for example as
\begin{equation}
\mathcal N^{\textrm{1,b}}_0(\epsilon_0) 
	= N_0 (\epsilon_0)
\label{DOS.region1.b}
\end{equation}
with
\begin{equation}
N_0(\epsilon) \equiv \frac{i \sqrt{\Gamma_-(\epsilon)}}{\pi r_- \gamma_x^2}
	\left[
		a_1 \Pi \left( \frac{r_+^2}{\beta_+^2} \left| \frac{r_+^2}{r_-^2} \right. \right)
		- a_2 \Pi \left( \frac{r_+^2}{\beta_-^2} \left| \frac{r_+^2}{r_-^2} \right. \right)
		- K \left( \frac{r_+^2}{r_-^2} \right)
	\right]
\label{N_0.defn}
\end{equation}
(we remind the reader that there is an implicit dependence on $\epsilon$ in $r_\pm(\epsilon)$ here).  We have also defined
\begin{equation}
a_{1,2} = \frac{\beta_{\pm}^2 + \beta_{\mp}^2 r_+^2 r_-^2 - (r_+^2 + r_-^2)}{\beta_+^2 - \beta_-^2}.
\label{a.12}
\end{equation}
(We could also write the result as $\mathcal N^{\textrm{1,b}}_0(\epsilon_0) = \frac{1}{2} - N_0 (-\epsilon_0)$ in which we have related the answer to that in zone (d) by symmetry).

(d)  $ \frac{\omega_z - \Omega}{2} < \epsilon_0 < \frac{\omega_z + \Omega}{2} $ (Upper sheet).
For the upper sheet, we can draw our contour $\tilde{\mathcal{C}}_1$ along the imaginary axis.  
The integrated DOS is given by
\begin{equation}
\mathcal N^{\textrm{1,d}}_0(\epsilon_0) = \frac{1}{2} + N_0 (\epsilon_0).
\label{DOS.region1.d}
\end{equation}
We show in App. \ref{app:DOS.calc} how to obtain this result by analytic manipulation of the integral in Eq. (\ref{dos.Q}).

\begin{figure}
\includegraphics[width=0.9\textwidth]{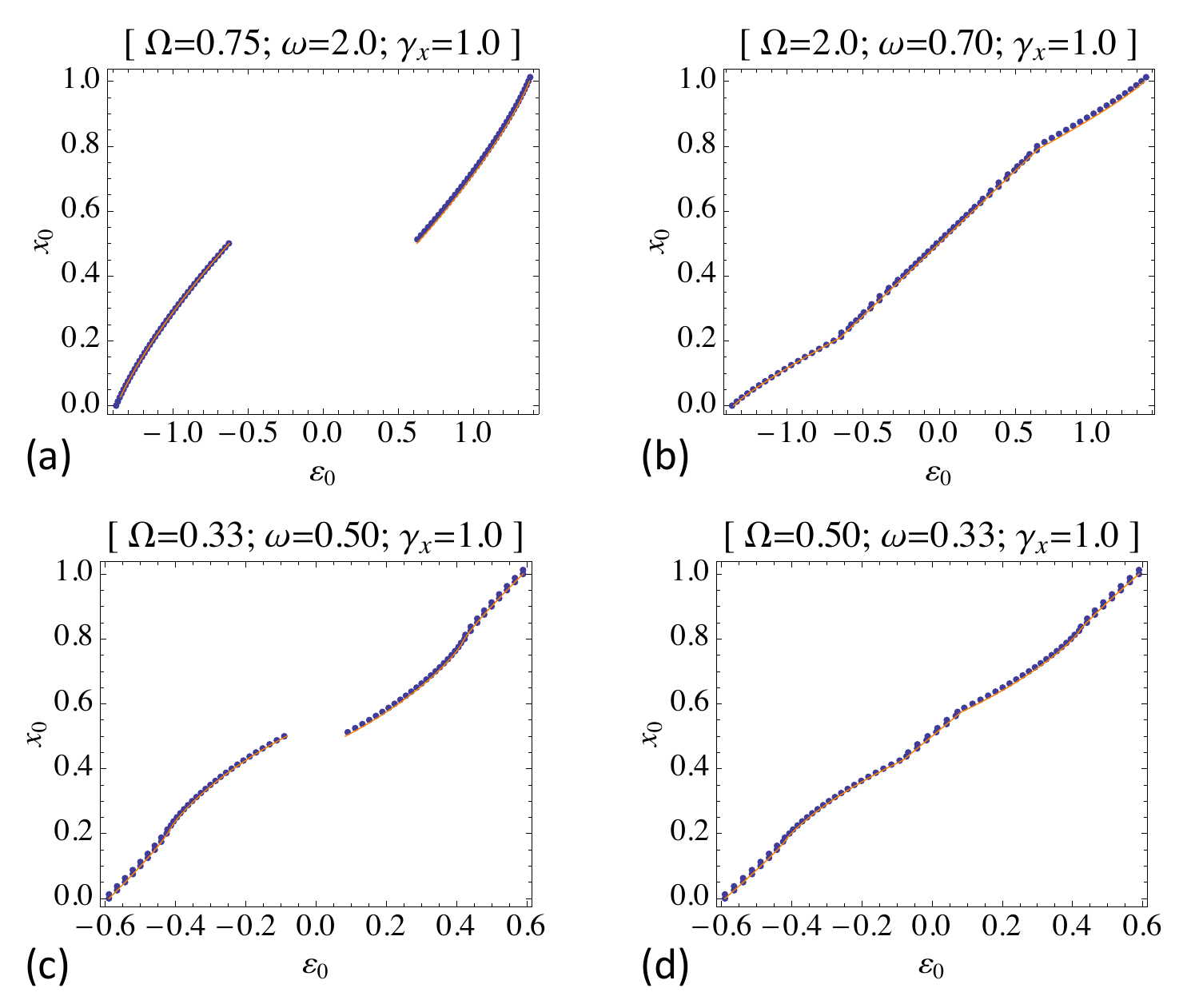}
\caption{(Color online) Analytic (orange curve) and numerical (blue dots)
integrated density of states in each region, with the parameter values as given above each plot (these are the same values as those used in Fig. \ref{fig:phase.regions} in each case).  
(a) Region 1, (b) Region 2, (c) Region 3, and (d) Region 4.
}
\label{fig:DOS.plots}
\end{figure}

We plot our analytic results for $\mathcal N^{\textrm{1}}_0(\epsilon_0) $ for the full spectrum as the orange curve in Fig. \ref{fig:DOS.plots}(a) against our numerical results for $\mathcal N^s (\epsilon)$ (blue dots). 
We show the results for Regions 2, 3 and 4 in the panels labeled (b), (c) and (d), respectively.

\subsubsection{Region 2 (symmetric, overlapping sheets)} \label{sec:anal.dos.expr.reg2}

Now we consider the case of Region 2 with non-degenerate ground state and overlapping energy surfaces.  This is given by $\omega_z > \gamma_x^2 / \Omega$ and $\omega_z < \Omega$ in Fig. \ref{fig:phase.regions}.  The results are similar to that for Region 1, but here there occurs zone (c) behavior in the overlapping region of the spectrum.

(b) $ - \frac{\omega_z + \Omega}{2} < \epsilon_0 < \frac{ \omega_z - \Omega}{2} $ (Non-overlapping portion of lower sheet).
For this portion of the spectrum in Region 2, we have simply $\mathcal N^{\textrm{2,b}}_0(\epsilon_0) = \mathcal N^{\textrm{1,b}}_0(\epsilon_0)$.

(c) $\frac{ \omega_z - \Omega}{2} < \epsilon_0 <  \frac{ - \omega_z + \Omega}{2} $ (Overlapping portion).  For the overlapping portion of the spectrum we have
\begin{equation}
\mathcal N^{\textrm{2,c}}_0(\epsilon_0) 
	=  \frac{1}{2} 
		- \left(\mathcal N^{\textrm{2,b}}_0(\epsilon_0) +  \mathcal N^{\textrm{2,d}}_0(\epsilon_0) \right)
\label{DOS.region2.c}
\end{equation}
where we give $N^{\textrm{2,d}}_0(\epsilon)$ immediately below.

(d) $ \frac{- \omega_z + \Omega}{2}  < \epsilon_0 < \frac{\omega_z + \Omega}{2}$ (Non-overlapping portion of upper sheet).
In the case of the upper sheet we have the results for $\mathcal N^{\textrm{2,d}}_0(\epsilon_0)$ as in zone I(d).

\subsubsection{Region 3 (broken phase, non-overlapping sheets)} \label{sec:anal.dos.expr.reg3}

Now we consider the case of region 3 with twice degenerate ground state (broken phase) and non-overlapping energy surfaces.  This is given by $\omega_z < \gamma_x^2 / \Omega$ and $\omega_z > \Omega$ in Fig. \ref{fig:phase.regions}.  Here the lower sheet is broken into zones (a) and (b), while the upper sheet is broken into (d) and (e).

(a) $- \frac{1}{2 \gamma_x}\sqrt{(\omega_z^2 + \gamma_x^2)(\Omega^2 + \gamma_x^2)} < \epsilon_0
		<  - \frac{\omega_z + \Omega}{2} $
	(Critical portion of lower sheet).
Throughout the degenerate portion of the lower sheet the contour $\mathcal{C}_1$ lies along two portions of the real axis as shown in Fig. \ref{fig:spheres.reg3.gs}.  This results from the two branch cuts overlapping.  In this case $r_+ > r_-$, so that the contour  $\mathcal{C}_1$ extends from $r_+$ to $r_-$ then from $-r_-$ to $-r_+$.
Although in practice, it is easier to work out the integration in terms of the second contour that extends from infinity back to the origin along the positive imaginary axis before returning to infinity along the negative imaginary axis.

We can obtain the integrated density of states in the critical region as
\begin{equation}
\mathcal N^{\textrm{3,a}}_0(\epsilon_0) 
	= \frac{1}{2} + N_0 (- \epsilon_0) - N_1 (- \epsilon_0)
\label{DOS.region3.a}
\end{equation}
in which $N_0 (\epsilon_0)$ is given in Eq. (\ref{N_0.defn}) and
\begin{equation}
N_1 (\epsilon) \equiv \frac{i \sqrt{\Gamma_-(\epsilon)}}{\pi r_+ \gamma_x^2}
	\left[
		a_2 \Pi \left( \frac{\beta_-^2}{r_+^2} \left| \frac{r_-^2}{r_+^2} \right. \right)
		- a_1 \Pi \left( \frac{\beta_+^2}{r_+^2} \left| \frac{r_-^2}{r_+^2} \right. \right)
		- r_+^2 r_-^2 \ K \left( \frac{r_-^2}{r_+^2} \right)
	\right].
\label{N_1.defn}
\end{equation}
We have obtained this result from symmetry in terms of that presented in zone 3(e) below, which can itself be obtained by following the method outlined in Appendix \ref{app:DOS.calc.crit}.

At the precise value $\epsilon_0 = - (\omega_z + \Omega)/2$ that gives the boundary between the critical and non-critical parts of the spectrum on the lower sheet (i.e. the DOS singularity),
we have $r_- = 0$ and therefore the integrated DOS given in the form of Eq. (\ref{dos.Lambda.form}) reduces to elementary integrals.  When computed, we find
\begin{equation}
\mathcal N^{\textrm{3}}_0(- \frac{\omega_z + \Omega}{2} ) 
	= \frac{\gamma_x \sqrt{\omega_z + \Omega} }{\pi \sqrt{\Omega (\Omega^2 + \gamma_x^2)} }
	\left[
	\sqrt{\beta_1^2 - r_+^2} \tan^{-1} \left( \frac{r_+}{\sqrt{\beta_1^2 - r_+^2}} \right)
	- \sqrt{\beta_2^2 - r_+^2} \tan^{-1} \left( \frac{r_+}{\sqrt{\beta_2^2 - r_+^2}} \right)
		\right].
\label{DOS.region3.critpt.lower}
\end{equation}

(b) $ - \frac{\omega_z + \Omega}{2} < \epsilon_0 < \frac{ - \omega_z + \Omega}{2} $ 
	(Non-critical portion of lower sheet).
For the non-critical portion of the lower sheet in Region 3, we have the usual result for 
$\mathcal N^{\textrm{3,b}}_0(\epsilon_0)$ the same as in Eq. (\ref{DOS.region1.b}).

(d) $ \frac{ \omega_z - \Omega}{2}  < \epsilon_0 < \frac{\omega_z + \Omega}{2}$
	(Non-critical portion of upper sheet).
In the case of the upper sheet we have the typical behavior for $\mathcal N^{\textrm{3,d}}_0(\epsilon_0)$
given in Eq. (\ref{DOS.region1.d}).

At the boundary $\epsilon_0 = (\omega_z + \Omega) / 2$ between the critical and non-critical parts of the spectrum on the upper sheet, the integrated DOS expression again simplifies.  In this case we have the result
\begin{equation}
\mathcal N^{\textrm{3}}_0(\frac{\omega_z + \Omega}{2} ) 
	= 1 - \mathcal N^{\textrm{3}}_0(- \frac{\omega_z + \Omega}{2} )
\label{DOS.region3.critpt.upper}
\end{equation}
in terms of Eq. (\ref{DOS.region3.critpt.lower}).

(e) $ \frac{\omega_z + \Omega}{2} < \epsilon_0 <
- \frac{1}{2 \gamma_x}\sqrt{(\omega_z^2 + \gamma_x^2)(\Omega^2 + \gamma_x^2)}$
(Critical portion of upper sheet).
In this case, we have the result
\begin{equation}
\mathcal N^{\textrm{3,e}}_0(\epsilon_0) 
	=  \frac{1}{2} - N_0 (\epsilon_0) + N_1 (\epsilon_0)
\label{DOS.region3.e}
\end{equation}
as outlined in the Appendix \ref{app:DOS.calc.crit}.

\subsubsection{Region 4 (broken phase, overlapping sheets)} \label{sec:anal.dos.expr.reg4}

Now we consider the case of Region 4 with degenerate ground state and overlapping energy surfaces.  This is given by $\omega_z < \gamma_x^2 / \Omega$ and $\omega_z < \Omega$ in Fig.
\ref{fig:phase.regions}.  Here the spectrum consists of zones (a) and (b) on the lower sheet, then an overlapping region (c) and then the remaining portion of the upper sheet consists of zones (d) and (e).

(a) $- \frac{1}{2 \gamma_x}\sqrt{(\omega_z^2 + \gamma_x^2)(\Omega^2 + \gamma_x^2)} < \epsilon_0
		<  - \frac{\omega_z + \Omega}{2} $
	(Critical portion of lower sheet).
Here the spectrum imitates exactly that from zone 3(a) with 
$\mathcal N^{\textrm{4,a}}_0(\epsilon_0) 
	= \mathcal N^{\textrm{3,a}}_0(\epsilon_0)$
as given in Eq. (\ref{DOS.region3.a}).

At the critical boundary $\epsilon_0 = - (\omega_z + \Omega)/2$ we again have 
$N^{\textrm{4}}_0(- \frac{\omega_z + \Omega}{2} ) = N^{\textrm{3}}_0(- \frac{\omega_z + \Omega}{2} )$ as given in Eq. (\ref{DOS.region3.critpt.lower}).

(b) $ - \frac{\omega_z + \Omega}{2} < \epsilon_0 < \frac{ \omega_z - \Omega}{2} $ 
	(Non-overlapping portion of non-critical lower sheet).
Here again we have the usual result for $\mathcal N^{\textrm{4,b}}_0(\epsilon_0)$ as given in Eq. (\ref{DOS.region1.b}).

(c) $\frac{ \omega_z - \Omega}{2} < \epsilon_0 <  \frac{ - \omega_z + \Omega}{2} $ (Overlapping portion).  For the overlapping spectrum we have
\begin{equation}
\mathcal N^{\textrm{4,c}}_0(\epsilon_0) 
	=  \frac{1}{2} 
		- \left(\mathcal N^{\textrm{4,b}}_0(\epsilon_0) +  \mathcal N^{\textrm{4,d}}_0(\epsilon_0) \right)
\label{DOS.region4.c}
\end{equation}
similar to the result given in zone 2(c) in Eq. (\ref{DOS.region2.c}).

(d) $ \frac{ - \omega_z + \Omega}{2}  < \epsilon_0 < \frac{\omega_z + \Omega}{2}$
	(Non-overlapping portion of non-critical upper sheet).
In the case of the non-overlapping, non-critical part of the upper sheet we have again the typical behavior for $\mathcal N^{\textrm{4,d}}_0(\epsilon_0)$
given in Eq. (\ref{DOS.region1.d}).

At the upper critical boundary $\epsilon_0 = (\omega_z + \Omega) / 2$ we again have
$\mathcal N^{\textrm{4}}_0(\frac{\omega_z + \Omega}{2} ) 
	= 1 - \mathcal N^{\textrm{4}}_0(- \frac{\omega_z + \Omega}{2} )$
similar to Eq. (\ref{DOS.region3.critpt.lower}) in Region 3.

(e) $ \frac{\omega_z + \Omega}{2} < \epsilon_0 <
- \frac{1}{2 \gamma_x}\sqrt{(\omega_z^2 + \gamma_x^2)(\Omega^2 + \gamma_x^2)}$
(Critical portion of upper sheet).
Here we again have
$\mathcal N^{\textrm{4,e}}_0(\epsilon_0) 
	= \mathcal N^{\textrm{3,e}}_0(\epsilon_0) $
as given in Eq. (\ref{DOS.region3.e}).


\section{Expectation values of system observables}\label{SEC:observables}

In this section we present the expectation values for several key observables in the Mermin model.  For the most important observables (those appearing in the Hamiltonian Eq. (\ref{ham.mermin})) we can easily obtain expectation values from our analytic results by applying the Hellmann-Feynman theorem.  The Hellmann-Feynman theorem relates the expectation values of observables to partial derivatives with respect to the system parameters as they appear in the energy eigenvalues.  In our case we easily obtain, for example,
\begin{equation}
\bra \sigma_z \ket = 2 \frac{\partial \epsilon}{\partial \omega_z} 
	= 2 \frac{\partial \epsilon}{\partial \mathcal N} \frac{\partial \mathcal N}{\partial \omega_z}
\label{exp.vals}
\end{equation}
where we have introduced partial derivatives in terms of the integrated density of states $\mathcal N$ since our expressions for this quantity given in Sec. \ref{sec:anal.dos.expr} are too complicated to invert analytically for $\epsilon$.

\begin{figure*}
\includegraphics[width=0.90\textwidth]{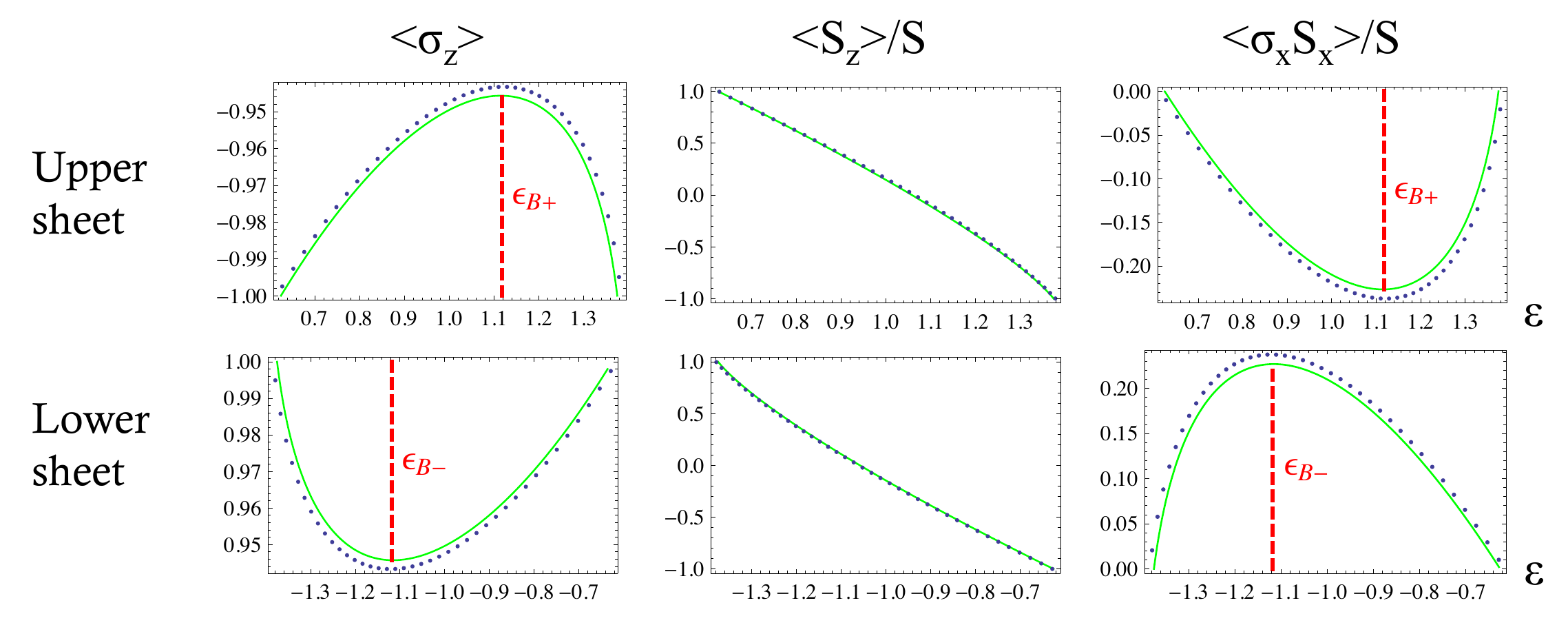}
\caption{(Color online) Comparison of expectation values for the observables appearing in the Hamiltonian Eq. (\ref{ham.mermin}) for Region 1.  Respectively by column we have plotted $\bra \sigma_z \ket$, $\bra S_z \ket$ and $\bra \sigma_x S_x \ket$ across the range of both energy sheets for the system parameter values $\omega_z = 2$, $\Omega=3/4$ and $\gamma_x = 1$.  The green curves represent the results from the Hellmann-Feynman theorem and the blue dots are numerical results computed assuming $N = 40$ as the number of background spins.  The red dashed lines indicate the position of the energy $\epsilon_{B\pm}$ as given in Eq. (\ref{epsilon.B.pm}).
}
\label{fig:exp.vals.R1}
\end{figure*}

In Fig. \ref{fig:exp.vals.R1} we plot the expectation values for $\bra \sigma_z \ket$, $\bra S_z \ket$ and $\bra \sigma_x S_x \ket$, respectively across the full spectrum for a representative case in Region 1 (symmetric case with non-overlapping sheets).  We have also included the numerical results, calculated directly from the Majorana polynomials using a system with $N = 40$ background spins.  In the lower left panel we have plotted the expectation value for the central spin $\bra \sigma_z \ket$ across the energy range of the lower sheet.  We see that at energy values $\epsilon$ near the ground state, this spin points upward, consistent with the picture presented in Fig. \ref{fig:sym.broken.spheres}(a).  Meanwhile, in the middle lower diagram the background spin $\bra S_z \ket$ also points up at energies near the ground state.  As we consider higher excitation states, we see that the $z$ component of both spins decreases as the effect of the coupling tends to pull the two spins towards anti-alignment on the $x$-axis.  However, the central spin reaches a minimum value partway across the range of the lower sheet and begins to re-align with the positive $z$-axis (the approximate energy value $\epsilon_{B-}$ where this occurs is labeled with a dashed red line).  Meanwhile, the $z$ component for the background spin continues leveling out, until it actually becomes negative in the upper half of the lower sheet.  Finally, at  the top of the lower sheet the two spins are anti-aligned in the $z$-axis.  This is consistent with the picture presented in the lower right-hand panel for $\bra \sigma_x S_x \ket$.  Here we see that the anti-alignment of the $x$-component of the spins is zero at the ground state, reaches a maximum partway across the energy sheet at $\epsilon_{B-}$ then returns to zero at the top of the lower energy sheet.

A similar behavior occurs in the range of the upper sheet, only with the central spin having a negative instead of a positive alignment on the $z$ axis.  Also $\bra \sigma_x S_x \ket \le 0$ in the upper sheet, which indicates that in this case alignment on the $x$-axis occurs between the two spins, instead of anti-alignment.

In theory, we could determine the spectruml extremal points that give the values of $\epsilon_{\pm B}$ using our expressions for the integrated density of states given in Eqs. (\ref{DOS.region1.b}) and (\ref{DOS.region1.d}), but the expressions are too complicated to invert them to explicitly obtain the spectrum.  However, we can still determine the approximate analytic position of $\epsilon_{\pm B}$ using our mean field theory expression for the energy given in Eq. (\ref{clas.energy.surface}).  We determine the extremal points in the expectation value for $\sigma_z$ according to
\begin{equation}
\frac{\partial}{\partial \alpha} \frac{\partial}{\partial \omega_z} \epsilon_\pm (\alpha, \bar{\alpha}) =
\frac{\partial}{\partial \bar{\alpha} } \frac{\partial}{\partial \omega_z} \epsilon_\pm (\alpha, \bar{\alpha}) 
	= 0.
\label{crit.cond.exp.vals}
\end{equation}
This condition yields the `breaking point' values for $\alpha, \bar{\alpha}$ according to $\alpha_B = \bar{\alpha}_B = \pm 1$.  Plugging these values back into Eq. (\ref{clas.energy.surface}) yields the breaking point energies as
\begin{equation}
\epsilon_{B\pm} = \pm \frac{ \sqrt{\omega_z^2 + \gamma_x^2} }{2}.
\label{epsilon.B.pm}
\end{equation}
Note that there is no dependence on $\Omega$ in this expression, only the coupling to the environment affects the position of these extremal points at this order of approximation.  These points are labeled by the red dashed lines appearing in Figs. \ref{fig:exp.vals.R1}.

\begin{figure*}
\includegraphics[width=0.90\textwidth]{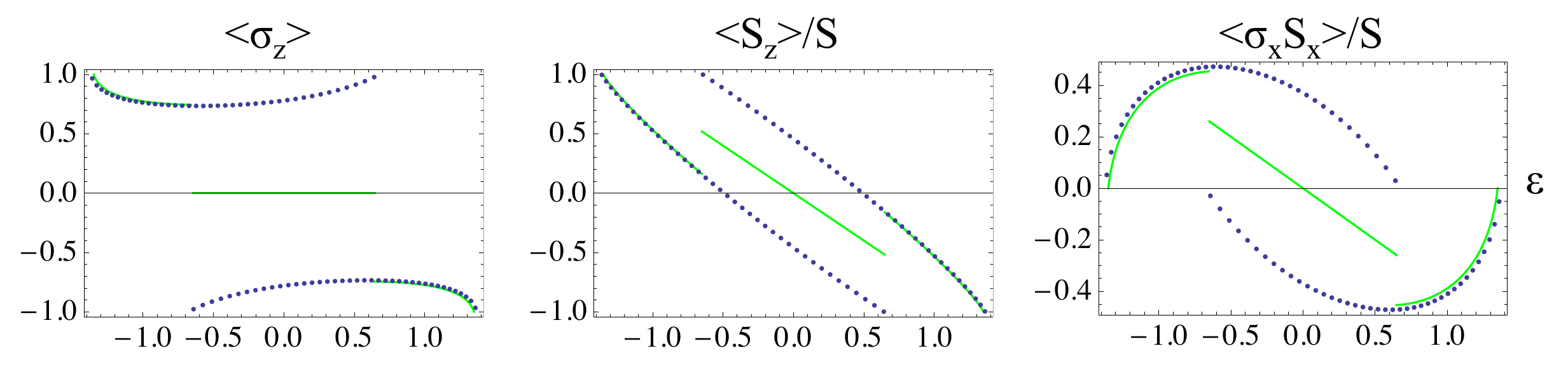}
\caption{(Color online) Comparison of expectation values for the observables appearing in the Hamiltonian Eq. (\ref{ham.mermin}) for Region 2.  Respectively by column we have plotted $\bra \sigma_z \ket$, $\bra S_z \ket$ and $\bra \sigma_x S_x \ket$ across the spectrum for the system parameter values $\omega_z = 7/10$, $\Omega=2$ and $\gamma_x = 1$.   The green curves represent the results from the Hellmann-Feynman theorem and the blue dots are numerical results computed assuming $N = 40$ as the number of background spins.
}
\label{fig:exp.vals.R2}
\end{figure*}

In Fig. \ref{fig:exp.vals.R2} we plot these same three expectation values for a case representative of Region 2, including numerical results.  Here each plot can be broken up into three sections along the lines of Sec. \ref{sec:anal.dos.expr.reg1}.  We see immediately that the numerical results give us two distinct ``bands,'' even in the overlapping region at the horizontal center of each plot.  This can be viewed in relation to Fig. \ref{fig:spheres.reg2.overlap}.  We have seen in this figure that increasing energies give an alternating `halo' pattern of zeros on the Majorana sphere in the overlapping region; similarly, we see in Fig. \ref{fig:exp.vals.R2} that increasing energies give states with expectation values that alternate between the two sheets in this region.

However, the analytic results obtained from application of the Hellmann-Feynman theorem (green curve) give an averaged value in the overlapping region, hence we would lose some information if we were to rely solely on Hellmann-Feynman.  This behavior is analogous to that occuring in Region IV of the LMG model as reported in \cite{Ribeiro1}.

\begin{figure*}
\includegraphics[width=0.90\textwidth]{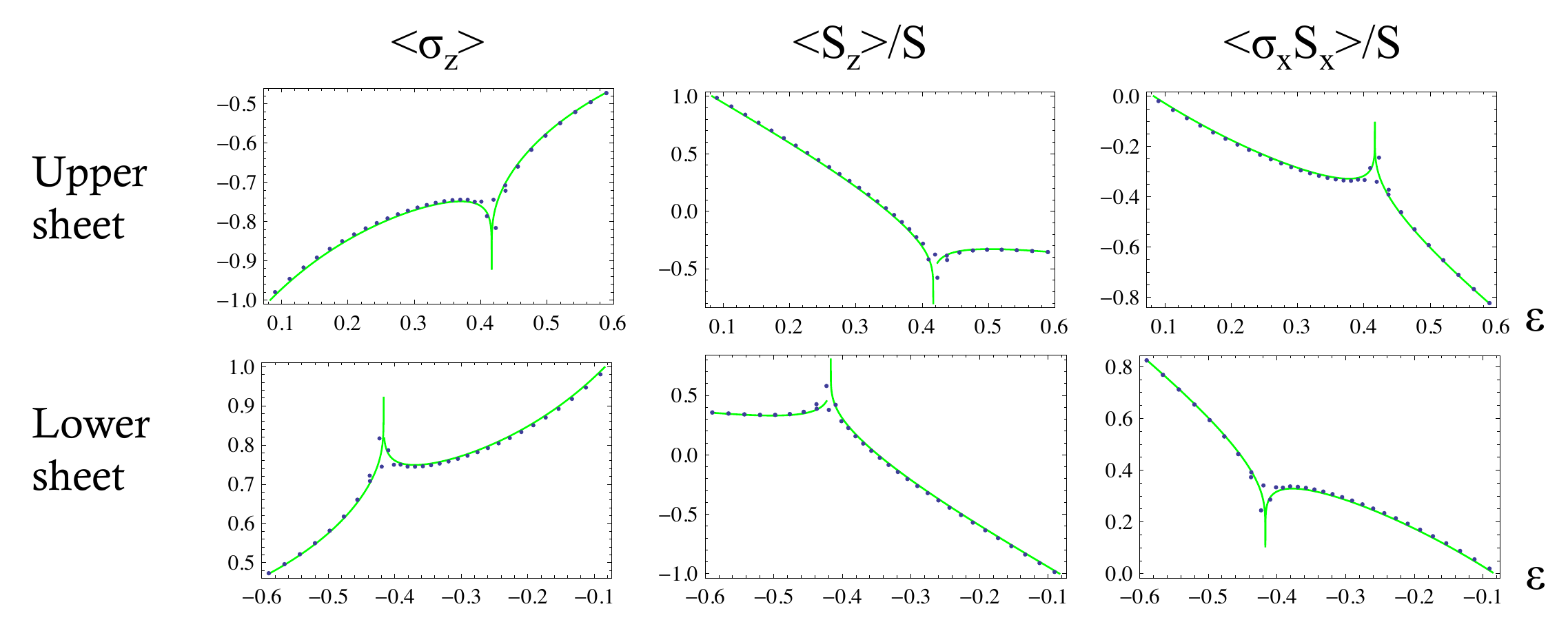}
\caption{(Color online) Comparison of expectation values for the observables appearing in the Hamiltonian Eq. (\ref{ham.mermin}) for Region 3.  Respectively by column we have plotted $\bra \sigma_z \ket$, $\bra S_z \ket$ and $\bra \sigma_x S_x \ket$ across the range of energies for both sheets, as labeled, for the system parameter values $\omega_z = 1/2$, $\Omega=1/3$ and $\gamma_x = 1$.  The green curves represent the results from the Hellmann-Feynman theorem and the blue dots are numerical results computed assuming $N = 40$ as the number of background spins.
}
\label{fig:exp.vals.R3}
\end{figure*}

We plot the same set of expectation values for a representative case of Region 3 in Fig. \ref{fig:exp.vals.R3} for $\omega_z = 1/2$, $\Omega=1/3$ and $\gamma_x = 1$.  We see that these expectation values are sharply peaked at the critical point at $\epsilon^0_- , \epsilon^\infty_+ = \mp 5/12 \approx \mp 0.416667$ between the broken and symmetric portions of the spectrum, as we should expect.

\begin{figure*}
\includegraphics[width=0.90\textwidth]{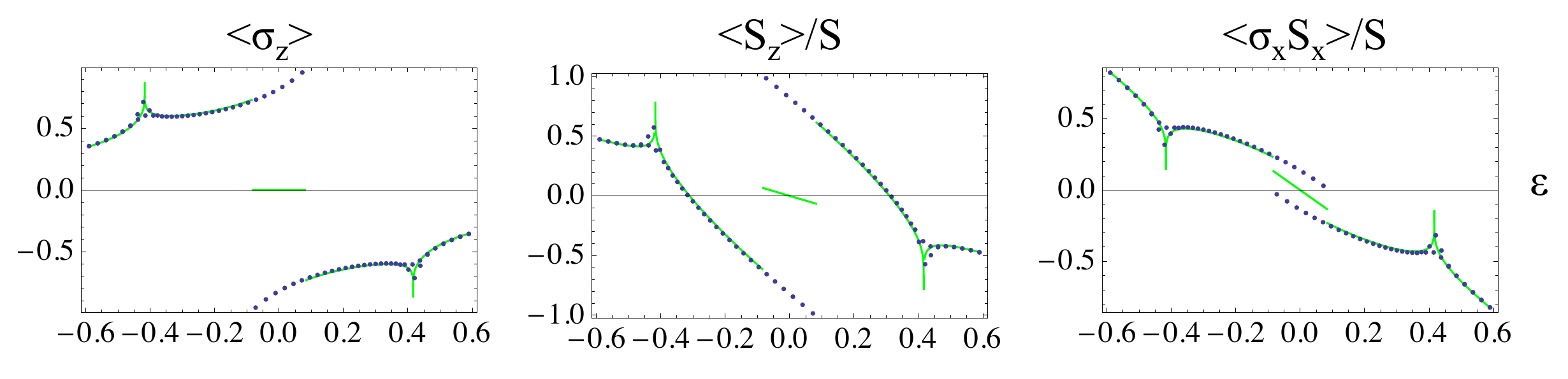}
\caption{(Color online) Comparison of expectation values for the observables appearing in the Hamiltonian Eq. (\ref{ham.mermin}) for Region 4.  Respectively by column we have plotted $\bra \sigma_z \ket$, $\bra S_z \ket$ and $\bra \sigma_x S_x \ket$ across the spectrum for the system parameter values $\omega_z = 1/3$, $\Omega=1/2$ and $\gamma_x = 1$.  The green curves represent the results from the Hellmann-Feynman theorem and the blue dots are numerical results computed assuming $N = 40$ as the number of background spins.
}
\label{fig:exp.vals.R4}
\end{figure*}

Finally we plot these expectation values for the system observables in Region 4 in Fig. \ref{fig:exp.vals.R4}.  Not surprisingly, we see a combination of the behaviors previously occurring in Regions 2 and 3.

\section{Conclusion}\label{SEC:conclusion}

In this study we have analyzed Mermin's special case of the central-spin model with symmetric coupling to environmental modes and monochromatic bath in the framework of a spin-coherent states formalism.  By applying a mean field theory approximation, we were able to obtain the classical energy surfaces for the double spectrum as presented in Eq. (\ref{clas.energy.surface}) and plotted for representative cases in Fig. \ref{fig:energy_surface_plots}; our results here are in agreement with studies by other researchers \cite{Mermin,Lev2000,Lev&Muth2001}.  We also obtained a complete phase picture for the model as presented in Figs. \ref{fig:phase.quads}, \ref{fig:phase.regions}, including the quantum phase transition $| \omega_z | = \gamma_x^2 / | \Omega |$ between Regions 1, 2 (symmetric phase) and Regions 3, 4 (broken phase) as well as the crossover point $| \omega_z | = | \Omega |$ for overlapping energy sheets between Regions 1, 3 (non-overlapping) and Regions 2, 4 (overlapping).

Moving beyond the mean field theory approach, from the time-independent Schr\"{o}dinger equation we were able to obtain the $G_0$ function in the thermodynamic limit in Eq. (\ref{G0}).  Relying on the integration contours we discovered from Figs. \ref{fig:spheres.reg1.gs} - \ref{fig:spheres.reg2.overlap}, we were able to write a closed form expression for the integrated density of states (for each phase region) in terms of an integration of the non-analytic part of the $G_0$ function over one of these contours.  This integral is presented in generic form in Eq. (\ref{dos.Q}).  We used this result in Sec. \ref{sec:anal.dos.expr} to work out the integrated density of states in analytic form for every zone of the spectrum in each of the four phase regions.  Some details of how to obtain these expressions by analytic manipulations of the integral in Eq. (\ref{dos.Q}) are presented in the Appendix.  Finally in Sec. \ref{SEC:observables} we plotted numerical results for the observables appearing in the Hamiltonian for representative cases in each of the four regions and applied the Hellmann-Feynman theorem to obtain an accurate comparison from our analytic results for the integrated density of states.

We envision several avenues for future study on the Mermin model as well as related models.
For example, in the Mermin model it might be interesting to consider in greater detail the $1/S$ corrections to the $G^{s,a}$ function in Eq. (\ref{G.expansion.sa}), as it is at first order in $1/S$ that the symmetric and anti-symmetric functions can be differentiated from one another according to Eqs. (\ref{P012.sa}).  While these calculations would be involved, they would allow us to determine the spectruml differences between the two sectors. 

Dynamical studies of the Mermin model may also provide the opportunity to study the evolution of a two-level system in the presence of a non-Markovian bath.  In particular, a semi-classical analysis building from the formalism presented in this paper may prove fruitful.  Other approaches to the dynamics of the Mermin model have already been explored in Refs. \cite{Ell&Kov} and \cite{Non-Markovian}.  The presence of non-Markovian effects has been demonstrated in \cite{Non-Markovian}.

One further well-posed problem would be to study the generalization of the present model to a central system with $s=1$, or even a two qubit central system for which both the internal entanglement as well as that between the central system and the bath could be computed.

\begin{acknowledgements}
S. G. would like to thank Xuedong Hu and Dvira Segal for stimulating discussion on the Mermin model, as well as support from CQIQC at the University of Toronto.
P. R. acknowledges support through FCT BPD grant SFRH/BPD/43400/2008.
\end{acknowledgements}

\appendix

\section{Explicit calculations related to the integrated density of states}\label{app:DOS.calc}

\subsection{Transformation of integrated density of states in non-critical zone (d)}\label{app:DOS.calc.non-crit}

In this section of the appendix we demonstrate how to manipulate the integral in Eq. (\ref{dos.Q}) to
obtain the result given in Eq. (\ref{DOS.region1.d}) for the (d) zone of the DOS.
Setting aside an $\epsilon_0$-dependent coefficient, the integral we should consider is given by
\begin{equation}
\int_0^{r_+}  \frac{ \sqrt{(\alpha^2 - r_+^2)(\alpha^2 - r_-^2)} }
								{(\alpha^2 - \beta_1^2)(\alpha^2 - \beta_2^2)} d \alpha.
\label{integral.zone.d}
\end{equation}
It is known that any integral over a rational function of $\alpha$ and
\begin{equation}
y(\alpha) \equiv \sqrt{(\alpha^2 - r_+^2)(\alpha^2 - r_-^2)}
\label{y.alpha}
\end{equation}
can be written as a combination of the three types of elliptic integrals in standard form
\cite{Lawden}
(in general $(y(\alpha))^2$ should be any cubic or quartic in $\alpha$ for this statement to apply).

To obtain the integral above in standard form, we assume that the integrand can be written as a sum of integrands in the form
\begin{equation}
\frac{y(\alpha)}{(\alpha^2 - \beta_1^2)(\alpha^2 - \beta_2^2)}
	= \frac{A}{y(\alpha)}
		+ \frac{B}{(\alpha^2 - \beta_1^2) y(\alpha)}
		+ \frac{C}{(\alpha^2 - \beta_2^2) y(\alpha)}
		+ D \; y(\alpha)
\label{integrand.expansion}
\end{equation}
in which $A, B, C$ and $D$ are undetermined constants.  After slight re-arrangement these terms would correlate to the integrands for the elliptic integral of the first kind, two possible elliptic integrals of the third kind and the elliptic integral of the second kind, respectively.  To determine these unknown constants we multiply Eq. (\ref{integrand.expansion}) through by a factor of $y(\alpha) (\alpha^2 - \beta_1^2)(\alpha^2 - \beta_2^2)$, yielding a quartic polynomial in $\alpha^2$ from which we can determine the unknowns $A, B, C$ and $D$.  We immediately obtain $D=0$ (no elliptic integral of the second kind) as well as $A = 1$, $B = \beta_+^2 a_1$ and $C = - \beta_-^2 a_2$, yielding the coefficients given in Eq. (\ref{a.12}).

From this point, re-arranging each integrand slightly and performing a transformation of the integration variable according to $q = \alpha / r_+$ allows us to obtain the result in Eq. (\ref{DOS.region1.d}) in terms of elliptic integrals in standard form.

\subsection{Integration form at the saddle-points in Regions 3 and 4}

In general, the integral form for the integrated DOS in Eq. (\ref{dos.Q}) can be re-written after a simple fraction decomposition as
\begin{equation}
\mathcal I_0 (\epsilon_0)
	= \frac{1}{i 2 \pi \Omega}
		\sqrt{\frac{\Gamma_- (\epsilon_0)}{\Omega^2 + \gamma_x^2}}
		\left[ \Lambda(\beta_1) - \Lambda(\beta_2)  \right]
\label{dos.Lambda.form}
\end{equation}
with the integral $\Lambda(\beta)$ given by
\begin{equation}
\Lambda (\beta) = \int_{C_1}
				\frac{\sqrt{ (\alpha^2 - r_+^2)(\alpha^2 - r_-^2)}}					
					{\alpha^2 - \beta^2} d \alpha.
\label{Lambda.defn}
\end{equation}
In the special case $\epsilon_0 = - (\omega_z + \Omega)/2$ in Regions 3 and 4 we have $r_- = 0$ and the integral $\Lambda (\beta)$ reduces so that it can be computed through elementary techniques, yielding Eq. (\ref{DOS.region3.critpt.lower}).

\subsection{Transformation of DOS integral in the critical zone (e)}\label{app:DOS.calc.crit}

In this region, it is easiest to re-write the integral in terms of a contour that extends from infinity back to the origin along the positive imaginary axis and then back out to infinity on the negative imaginary axis.  We can re-write the integration variable in Eq. (\ref{dos.Q}) along the imaginary axis according to $i q = \alpha$.  Noting the symmetry properties of the resulting integral, it can be written as proportional to
\begin{equation}
Y_0(\epsilon) \equiv \frac{\sqrt{\Gamma_-(\epsilon)}}{\pi \gamma_x^2}
	\int_{0}^\infty \frac{\sqrt{\left(q^2 + r_+^2\right)\left(q^2 + r_-^2\right)} }
					{\left( 1 + q^2 \right)^2 + \frac{4 \Omega^2}{\gamma_x^2}q^2 } dq.
\label{Y0.defn}
\end{equation}
We can break $Y_0 (\epsilon)$ into two integrals, one over $(0,r_+)$ and a second over $(r_+,\infty)$.  The first integral (re-including all appropriate coefficients) yields $N_0$ from Eq. (\ref{N_0.defn}).

For the remaining integral on the interval $(r_+,\infty)$ we perform a simple inversion of the integration variable according to $p^{-1} = q$.  After some slight re-arranging of the resulting integrand, we obtain a new integral along $(0, 1/r_+)$ that can again be calculated by following a method similar to that outlined in App. \ref{app:DOS.calc.non-crit} above.  After re-introducing coefficients, we obtain the result $N_1(\epsilon)$ as given in Eq. (\ref{N_1.defn}) in the main text.


\begin{thebibliography}{99} 


\bibitem{cent_spin_review}
N. V. ProkofÕev and P. C. E. Stamp,
Rep. Prog. Phys. {\bf 63}, 669 (2000)

\bibitem{spin_boson_review}
A. J. Leggett, S. Chakravarty, A. T. Dorsey, M. P. A. Fisher, A. Garg, and W. Zwerger,
Rev. Mod. Phys. {\bf 59}, 1 (1987)

\bibitem{UWeiss}
U. Weiss,
{\it Quantum Dissipative Systems}, 2nd ed.,
World Scientific, Singapore (1998).

\bibitem{Wubs2006}
Martijn Wubs, Keiji Saito, Sigmund Kohler, Peter H\"{a}nggi, and Yosuke Kayanuma,
Phys. Rev. Lett. {\bf 97}, 200404 (2006).




\bibitem{Mermin}
N. David Mermin,
Phys. A {\bf 177}, 561 (1991).

\bibitem{Lev2000}
Gregory Levine,
Phys. Rev. B, {\bf 61}, 4636 (2000).

\bibitem{Lev&Muth2001}
Gregory Levine and V. N. Muthukumar,
Phys. Rev. B {\bf 63}, 245112 (2001).

\bibitem{Ell&Kov}
D. Ellinas and V. Kovanis,
Phys. Rev A {\bf 51} 4230 (1995).

\bibitem{Non-Markovian}
Heinz-Peter Breuer, Daniel Burgarth, and Francesco Petruccione,
Phys. Rev. B {\bf 70}, 045323 (2004).




\bibitem{Raddiffe}
J. M. Raddiffe,
J. Phys. A {\bf A}, 313 (1971).

\bibitem{Arrechi}
F. T. Arrecchi, E. Courtens, R. Gilmore, and H. Thomas,
Phys. Rev. A {\bf 6}, 2211 (1971).

\bibitem{Klauder}
J. R. Klauder and B. S. Skagerstam,
{\it Coherent States}, World Scientific, Singapore (1985);
Askold Perelomov,
{\it Generalized Coherent States and Their Applications},
Springer-Verlag New York (1986).



\bibitem{Majorana}
E. Majorana,
Nuovo Cimento {\bf 9}, 43 (1932).




\bibitem{Ribeiro1}
Pedro Ribeiro, Julien Vidal, and R\'{e}my Mosseri,
Phys. Rev. Lett. {\bf 99}, 050402 (2007).

\bibitem{Ribeiro2}
Pedro Ribeiro, Julien Vidal, and R\'{e}my Mosseri,
Phys. Rev. E 78, 021106 (2008).



\bibitem{Fried&Dirk1}
F. Trimborn, D. Witthaut, and H. J. Korsch,
Phys. Rev. A {\bf 77}, 043631 (2008).


\bibitem{Huang}
Kerson Huang
{\it Statistical Mechanics}, 2nd Ed., John Wiley \& Sons (1987).

\bibitem{Mosseri&Dand}
R\'emy Mosseri and Rossen Dandoloff,
J. Phys. A: Math. Gen. {\bf 34}, 10243 (2001).

\bibitem{Lawden}
Derek F. Lawden, {\it Elliptic Functions and Applications},
Springer-Verlag New York (1989).

\end{thebibliography}
\end{document}